\documentclass[12pt]{article}
\pdfoutput=1

\usepackage{feynmp}
\usepackage{amsmath,amssymb,amssymb,graphicx,bbm} 
\usepackage{cite}
\usepackage{hyperref}
\usepackage{slashed}
\usepackage{subfigure}

\newcommand{\beq}{\begin{eqnarray}}
\newcommand{\eeq}{\end{eqnarray}}

\numberwithin{equation}{section}

\newcommand{\centeron}[2]{{\setbox0=\hbox{#1}\setbox1=\hbox{#2}\ifdim
\wd1>\wd0\kern.5\wd1\kern-.5\wd0\fi \copy0
\kern-.5\wd0\kern-.5\wd1\copy1\ifdim\wd0>\wd1
                                   \kern.5\wd0\kern-.5\wd1\fi}}
\newcommand{\ltap}{\>\centeron{\raise.35ex\hbox{$<$}}
                           {\lower.65ex\hbox{$\sim$}}\>}
\newcommand{\gtap}{\>\centeron{\raise.35ex\hbox{$>$}}
                           {\lower.65ex\hbox{$\sim$}}\>}
\newcommand{\gsim}{\mathrel{\gtap}}
\newcommand{\lsim}{\mathrel{\ltap}}

\newcommand\ZZ{\hbox{\zfont Z\kern-.4emZ}}
\font\zfont = cmss10 

\newcommand{\drawsquare}[2]{\hbox{%
\rule{#2pt}{#1pt}\hskip-#2pt
\rule{#1pt}{#2pt}\hskip-#1pt
\rule[#1pt]{#1pt}{#2pt}}\rule[#1pt]{#2pt}{#2pt}\hskip-#2pt
\rule{#2pt}{#1pt}}

\newcommand{\fund}{\ensuremath{\drawsquare{6.5}{0.4}}}

\newcommand{\symm}{\ensuremath{\drawsquare{6.5}{0.4}\hskip-0.4pt%
        \drawsquare{6.5}{0.4}}}

\newcommand{\sing}{\ensuremath{\mathbf{1}}}

\newcommand{\id}{\ensuremath{\mathbf{1}}}

\def\lpar#1#2#3#4{\rlap{\raise#3\hbox{$\hskip#4#1\left\{\mbox{\phantom{\rule[0mm]{0mm}{#2}}}\right.$}}}
\def\rpar#1#2#3#4{\rlap{\raise#3\hbox{$\hskip#4\left\}#1\mbox{\phantom{\rule[0mm]{0mm}{#2}}}\right.$}}}

\newcommand{\bZ}{\mathbb{Z}}

\newcommand{\cM}{\mathcal{M}}

\newcommand{\ov}{\overline}

\newcommand{\SU}{\mathrm{SU}}

\newcommand{\U}{\mathrm{U}}

\newcommand{\rd}[1]{\mathbf{#1}}

\begin{document}
\setlength{\unitlength}{1mm}
\begin{titlepage}

\vskip.5cm
\begin{center}
{\Large \bf A Complete Model for R-parity Violation
 }

\vskip.1cm
\end{center}
\vskip0.2cm

\begin{center}
{
{Csaba Cs\'aki} 
{\rm and}
{Ben Heidenreich}}
\end{center}
\vskip 8pt

\begin{center}
{\it  Department of Physics, LEPP, 
Cornell University, Ithaca, NY 14853} \\
\vspace*{0.1cm}
 
\vspace*{0.3cm}
{\tt  csaki,bjh77@cornell.edu}
\end{center}

\vglue 0.3truecm

\begin{abstract}
\vskip 3pt

\noindent We present a complete model whose low energy effective theory is the R-parity violating NMSSM with a baryon number violating $\bar{u}\bar{d}\bar{d}$ vertex of the MFV SUSY form, leading to prompt LSP decay and evading the ever stronger LHC bounds on low-scale R-parity conserving supersymmetry. MFV flavor structure is enforced by  gauging an $\SU(3)$ flavor symmetry at high energies. After the flavor group is spontaneously broken, mass mixing between the standard model fields and heavy vector-like quarks and leptons induces hierarchical Yukawa couplings which depend on the mixing angles. The same mechanism generates the $\bar{u}\bar{d}\bar{d}$ coupling, explaining its shared structure. A discrete R-symmetry is imposed which forbids all other dangerous lepton and baryon-number violating operators (including Planck-suppressed operators) and simultaneously solves the $\mu$ problem. While flavor constraints require the flavor gauge bosons to be outside of the reach of the LHC, the vector-like top partners could lie below 1 TeV.

\end{abstract}

\end{titlepage}


\section{Introduction}
\label{sec:intro}
\setcounter{equation}{0}
\setcounter{footnote}{0}

Supersymmetry (SUSY) broken at the TeV scale has long been considered the leading candidate for a solution to the hierarchy problem of the Standard Model (SM).
 However, the first two years of LHC data do not contain any hints of the traditional signals of SUSY~\cite{LHCSUSY}, pushing the superpartner mass scale to uncomfortably high values in the simplest implementations of the theory, too high to solve the hierarchy problem without introducing other tunings. The recent discovery~\cite{higgsdiscovery} of the Higgs boson at around 126 GeV puts additional pressure on minimal SUSY: it is quite difficult to achieve such a heavy Higgs mass within the simplest models without tuning~\cite{higgsnaturalness}. If low-scale SUSY is nonetheless realized in nature, it is likely that one or more additional ingredients beyond the minimal version are present.

There are several known ways to avoid the direct superpartner searches, including raising the mass of the first two generation squarks and the gluino~\cite{moreminimal,naturalsusy} (``natural SUSY"), a compressed or stealthy spectrum~\cite{stealth}, an R-symmetric theory with Dirac gaugino masses~\cite{Rsymmetric}, and R-parity violation~\cite{MFVSUSY,RPVLHC}. Similarly, the Higgs mass can be raised by extending the theory to the NMSSM, possibly by making the Higgs and the singlet composite~\cite{compositehiggs}, or by strengthening the Higgs quartic interaction by introducing additional gauge interactions~\cite{gaugeextended}. In this paper we focus on the scenario where the lightest superpartner (LSP) decays promptly via an R-parity violating (RPV) vertex, evading the bounds from direct superpartner searches. We then introduce an NMSSM singlet to raise the Higgs mass to the required 126 GeV value.

It has long been known that RPV~\cite{RPV} can significantly change the collider phenomenology of SUSY models without leading to excessive baryon (B) and lepton (L) number violation (for a review see~\cite{RPVreview}). This is most easily accomplished in models where either B or L is conserved to a very good approximation, since the most stringent constraints on these couplings arise from the nonobservation of proton decay, which generally requires both B and L to be violated. The remaining couplings are subject to the relatively weaker constraints on processes which only violate B or L individually, and can be large enough to have a substantial impact on collider signatures. A particularly interesting possibility is when the LSP is a third generation (stop or sbottom) squark, decaying via the RPV operator $\bar{u}\bar{d}\bar{d}$ as $\tilde{t}\to \bar{b}\bar{s}$ or $\tilde{b}\to \bar{t}\bar{s}$, which is very difficult to disentangle from the vast amount of QCD background at the LHC~\cite{MFVSUSY,mesinooscillation}.\footnote{However, the gluino must be relatively heavy even in models with RPV, as decays to same sign tops will put a lower bound of order 700 GeV on the gluino mass, see for example~\cite{gluinobound}.}

One of the principle objections to RPV models is aesthetic in nature: one needs to introduce a large number of additional small parameters, which, while technically natural, is usually not very appealing. One possible simplifying assumption is to employ the hypothesis of minimal flavor violation (MFV)~\cite{MFV}. In MFV models the only sources of flavor violation are the SM Yukawa couplings. If one applies this hypothesis~\cite{Smith,MFVSUSY} to the SUSY SM one obtains a robust prediction~\cite{MFVSUSY} for the baryon-number violating RPV couplings: they will be related to the ordinary Yukawa couplings. Thus the BNV couplings for third generation quarks will be the largest, while those involving only light generations will be very strongly suppressed. The resulting simple model evades most direct LHC bounds while preserving naturalness of the Higgs mass, whereas the 126 GeV Higgs mass can be achieved by extending the model to the NMSSM. 

However, MFV is only 
a spurion counting prescription, rather than a full-blown effective theory. It does not fix the overall coefficients of the RPV terms, and does not even fix the relative coefficients of the baryon number violating (BNV) and lepton number  violating  (LNV) operators. Moreover, it is not obvious a priori that a complete theory can be formulated that produces MFV SUSY as its low-energy effective theory and ensures that LNV operators are sufficiently suppressed to avoid proton decay. The aim of this paper is to present a complete model that produces Yukawa-suppressed RPV terms in the low-energy effective theory. Since we want to explain the MFV structure of the entire effective Lagrangian, we will have to incorporate a full-fledged theory of flavor into the model. We assume that the flavor hierarchy arises due to (small) mixing with heavy vectorlike quarks and leptons. Upon integrating out these heavy fields, we obtain the SM flavor hierarchy as well as the Yukawa suppressed RPV terms. To ensure that only the operators compatible with MFV are generated, we will gauge an SU(3) subgroup of the SU(3)$^5$ spurious flavor symmetry of the standard model and impose a discrete symmetry to forbid other dangerous baryon and lepton number violating operators. 

The paper is organized as follows: in section~\ref{sec:blocks} we first review how to obtain flavor hierarchies from mixing with heavy flavors. We then describe an anomaly-free gauged $\SU(3)$ flavor symmetry which incorporates the heavy flavors, together with the flavor Higgs sector needed to spontaneously break this symmetry and introduce the required mass mixings to generate the SM Yukawa couplings.
 In section~\ref{sec:dangerousLNV} we analyze all gauge-invariant operators that can lead to excessive baryon and lepton number violation, deriving experimental constraints on their couplings to determine which operators must be forbidden by a discrete symmetry. In section~\ref{sec:completemodel} we present an anomaly-free discrete symmetry which forbids all problematic operators and describe the allowed flavor Higgs potential, completing the model. In section~\ref{sec:SUSYbreaking} we consider the structure of the induced soft SUSY breaking terms and comment on the possibility that the third generation of heavy vector-like quarks could be within the range of the LHC. We conclude in section~\ref{sec:conclusions}, presenting the details of our choice of a suitable anomaly-free discrete symmetry in an appendix. 

\section{The building blocks of the UV completed MFV SUSY}
\label{sec:blocks}
\setcounter{equation}{0}
\setcounter{footnote}{0}

The MFV SUSY scenario, outlined in~\cite{MFVSUSY}, is an R-parity violating
variant of the MSSM, with the superpotential
\begin{eqnarray}
  W & = & \mu H_u H_d + q Y_u  \bar{u} H_u + q Y_d  \bar{d} H_d + \ell Y_e 
  \bar{e} H_d + \frac{1}{2} w''  ( Y_u  \bar{u}) ( Y_d  \bar{d}) ( Y_d 
  \bar{d})
\label{eq:MFVSUSY}
\end{eqnarray}
and soft-terms with a minimal flavor-violationg (MFV) structure. The Yukawa
couplings, $Y_u$, $Y_d$, and $Y_e$, are holomorphic spurions charged under the
$\SU(3)_q \times \SU(3)_{\bar{u}} \times \SU(3)_{\bar{d}} \times \SU(3)_{\ell} \times \SU(3)_{\bar{e}}$ flavor
symmetry. Unlike ordinary R-parity conserving MFV, MFV SUSY imposes relations between
different \textit{superpotential} couplings, and there is no RG mechanism
for generating these relations, since the superpotential is not renormalized.
Thus, to explain the form of the superpotential beyond the level of a spurion
analysis, it is necessary to embed MFV SUSY within a high-scale model which
naturally generates this flavor structure.

Another reason that MFV SUSY requires a UV completion is that, while the superpotential~(\ref{eq:MFVSUSY}) is technically natural, it is not safe from Planck-suppressed corrections. For instance, the operator $\frac{1}{M_{\rm pl}} q^3 \ell$ may be generated by gravitational effects, whereas without an MFV structure this operator leads to rapid proton decay, as we show in section \ref{subsec:LNVconstraints}. Since global and/or spurious symmetries are generically broken by gravitational effects, to forbid this kind of operator we will ultimately require some additional gauge symmetry.
 
\subsection{Yukawa hierarchies from mixing with heavy matter}

One possibility would be to try to promote the entire (semi-simple) SM flavor symmetry $\SU(3)_q \times \SU(3)_{\bar{u}} \times \SU(3)_{\bar{d}}
\times \SU(3)_{\ell} \times \SU(3)_{\bar{e}}$ to a gauge symmetry, with the Yukawa couplings
arising as vevs of superfields. However, in this case, the superpotential
becomes nonrenormalizable, and in particular, the term
\begin{eqnarray}
  W & = & \frac{1}{\Lambda} q \Phi_u  \bar{u} H_u 
\end{eqnarray}
requires $\Phi_u$ to get a vev of the same order as the cutoff, due to the
$\mathcal{O} ( 1)$ top Yukawa coupling. The resulting effective field theory will necessarily have a low cutoff and will need its own UV completion. This suggests that we must introduce additional massive matter fields, which
generate the Yukawa couplings upon being integrated out. If the BNV couplings are generated along with the ordinary Yukawa couplings upon integrating out the heavy fields then this explains their related structure.

As an example consider a quark sector consisting of the usual light quarks $q,\bar{u},\bar{d}$ together with three pairs of vector-like right-handed up and down quarks $U,\bar{U}$ and $D,\bar{D}$, where $\bar{U}$ and $\bar{D}$ share the same SM quantum numbers as $\bar{u}$ and $\bar{d}$ respectively. We assume the superpotential
\begin{equation}
  W = \lambda_u q  \bar{U} H_u + \lambda_d q \bar{D} H_d + \frac{1}{2}
  \lambda_{\mathrm{bnv}}  \bar{U}  \bar{D}  \bar{D} +U \mathcal{M}_u \bar{U}
  +D \mathcal{M}_d \bar{D}  + U \mu_u \bar{u} + D \mu_d \bar{d} \,,
\label{eq:quarkW}
\end{equation}
where $\lambda_{u,d}$ and $\lambda_{bnv}$ are flavor-universal parameters while $\mathcal{M}_{u,d}$ and $\mu_{u,d}$ are in general $3\times3$ mass matrices. 
For $\mathcal{M} \gg \mu$, the low-energy effective theory will contain small effective Yukawa couplings for the chiral fields {\it and} an effective $\bar{u}\bar{d}\bar{d}$ BNV operator due to the mixing between $\bar{u}$ and $\bar{U}$ and between $\bar{d}$ and $\bar{D}$. At tree-level, one can integrate out the heavy fields using the $U$ and $D$ F-term conditions:
\begin{equation}
\bar{U} = - \mathcal{M}_u^{-1} \mu_u \bar{u}\,, \;\;\; \bar{D} = - \mathcal{M}_d^{-1} \mu_d \bar{d}\,,
\end{equation}
leading to the MFV SUSY superpotential~(\ref{eq:MFVSUSY}) with $w''=\lambda_{bnv}/(\lambda_u \lambda_d^2)$ and the Yukawa couplings
\begin{equation}
Y_x = \lambda_x \Upsilon_x \left( \id + \Upsilon_x^{\dag} \Upsilon_x \right)^{-1/2} \,, \;\;\; \Upsilon_x \equiv -\mathcal{M}_x^{-1} \mu_x \,, \label{eq:effYukawa}
\end{equation}
for $x=u,d$.\footnote{The factor in parentheses arises upon canonically normalizing the K\"{a}hler potential after integrating out the heavy fields.} This expression is readily understood by diagonalizing $\Upsilon_x$. Each eigenvalue\footnote{More precisely singular value.} $\sigma_i$ of $\Upsilon_x$ corresponds to the tangent of the corresponding mixing angle between the SM field $\bar{u}$ or $\bar{d}$ and the vectorlike partner $\bar{U}$ or $\bar{D}$. Since $\bar{U}$ and $\bar{D}$  couple directly to the Higgs with universal coupling $\lambda_{u,d}$, a small eigenvalue $\sigma_i \ll 1$ of $\Upsilon_x$ corresponds to a small Yukawa coupling $\lambda_x \sigma_i$, whereas a large eigenvalue $\sigma_i \gg 1$ of $\Upsilon_x$ corresponds to a maximal Yukawa coupling $\lambda_x$, with a smooth transition between the two behaviors around $\sigma_i \sim \mathcal{O}(1)$.

We see that hierarchical Yukawa couplings can arise if the mass matrices $\mathcal{M}$ and/or $\mu$ have hierarchical eigenvalues, whereas $w''$ is order one so long as the flavor universal couplings $\lambda_{u,d}$ and $\lambda_{bnv}$ are also order one. While other choices are possible, for the remainder of this paper we will assume for simplicity that $\mu_{u,d}$ are flavor-universal parameters, so that all the flavor structure is generated by $\mathcal{M}_{u,d}$. This choice is motivated by the possibility of observable collider signatures, as it allows the vector-like third-generation partners to be relatively light, since the mass matrix for the vector-like generations takes the form:
\begin{equation}
M^2_x = \mathcal{M}_x \mathcal{M}_x^{\dag} + \mu_x \mu_x^{\dag} = |\mu_x \lambda_x|^2 \left[Y_x Y_x^{\dag} \right]^{-1} \,,
\end{equation}
where the second equality follows in the case that $\mu_x$ is flavor-universal.

\begin{figure}
\begin{center}
\includegraphics[width=3in]{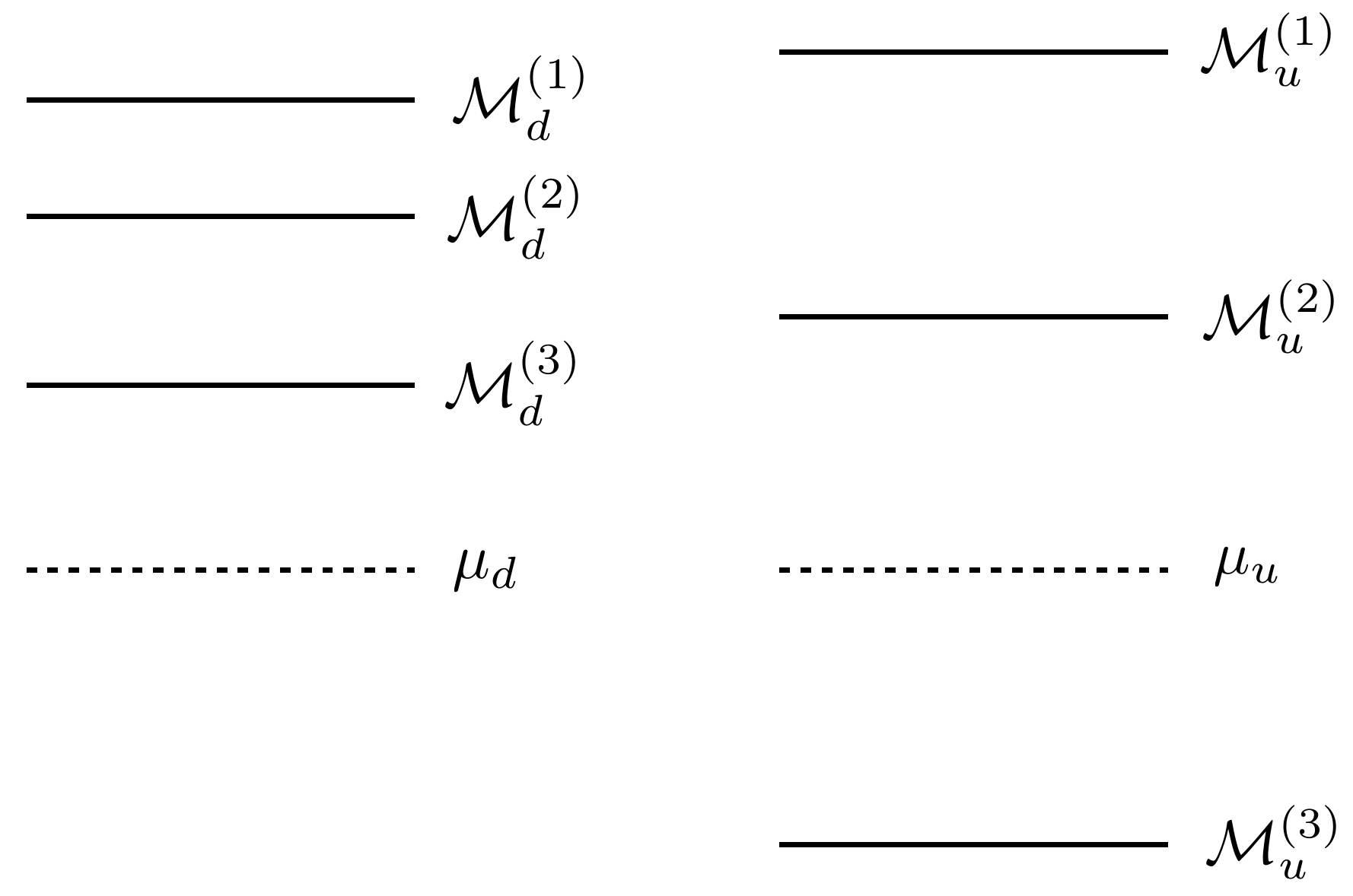}
\end{center}
\caption{A schematic illustration of the relative scales of the eigenvalues of $\mathcal{M}$ vs. $\mu$ for down-type (left) and up-type (right) quarks for $\lambda_{u,d}\sim \tan\beta \sim 1$. When $\mu > \mathcal{M}$ the Yukawa coupling will be unsuppressed, while all other Yukawas are suppressed by a factor of $\mu /\mathcal{M}$.\label{fig:topYukawa}}
\end{figure}

If $\lambda_{u,d} \lsim 1$, then $\mathcal{M}\gg \mu$ will generate only small Yukawa couplings. In order to accommodate the $\mathcal{O}(1)$ top Yukawa coupling, one eigenvalue of $\mathcal{M}_u$, which we denote $\mathcal{M}_u^{(3)}$, should be smaller than $\mu_u$. In this case one integrates out the fields $U^{(3)}$ and $\bar{u}^{(3)}$ at the scale $\mu$, and $\bar{U}^{(3)}$ will remain in the spectrum with a Yukawa coupling of order $\lambda_u$, as discussed above. The mass scales in~(\ref{eq:quarkW}) implied by the observed Yukawa couplings are schematically illustrated in Fig.~\ref{fig:topYukawa} for the case $\lambda_{u,d}\sim\tan\beta\sim1$.

A similar construction for the lepton sector (with SM fields denoted by $\ell,\bar{e}$) has several possible variants, yielding somewhat different expressions for the neutrino masses. One possibility involves a set of three heavy vectorlike RH charged leptons $E,\bar{E}$ and three RH neutrinos $\bar{N}$ with the superpotential 
\begin{equation}
W=\lambda_e \ell \bar{E} H_d +\lambda_n \ell \bar{N} H_u + E \mathcal{M}_e \bar{E} +\frac{1}{2} \bar{N} \mathcal{M}_n \bar{N} + E \mu_e \bar{e}\ ,
\label{eq:leptonW}
\end{equation}
which after integrating out the heavy fields yields just the SM Yukawa terms 
\begin{equation}
W_{\rm eff} = \ell Y_e \bar{e} H_d -\frac{1}{2} \lambda_n^2 (\ell H_u) \mathcal{M}_n^{-1} (\ell H_u) \ ,
\end{equation}
with $Y_e$ given by~(\ref{eq:effYukawa}).

Another possibility is to instead introduce three heavy lepton doublets $L,\bar{L}$ along with three RH neutrinos $\bar{n}$ and the superpotential 
\begin{equation}
W = \lambda_e L \bar{e} H_d +\lambda_n L \bar{n} H_u +L \mathcal{M}_\ell \bar{L} +\frac{1}{2} \bar{n} \mathcal{M}_n \bar{n} + \ell \mu_\ell \bar{L}\ ,
\end{equation}
which gives rise to the effective superpotential:
\begin{equation}
W_{\rm eff} = \ell Y_e \bar{e} H_d -\frac{1}{2} \frac{\lambda_n^2}{\lambda_e^2} (\ell H_u) Y_e \mathcal{M}_n^{-1} Y_e^T (\ell H_u)\, .
\label{eq:lepton2W}
\end{equation}
after integrating out the heavy fields, where now
\begin{equation}
Y_e = \lambda_e \left( \id + \Upsilon_{\ell} \Upsilon_{\ell}^{\dag} \right)^{-1/2} \Upsilon_{\ell} \,, \;\;\; \Upsilon_{\ell} \equiv -\mu_{\ell} \mathcal{M}_{\ell}^{-1}  \,.
\end{equation}

A third possibility, resulting in Dirac neutrino masses, is to introduce light RH neutrinos $\bar{n}$ together with vectorlike pairs or RH charged leptons $E,\bar{E}$ and neutrinos $N, \bar{N}$. We then impose lepton number conservation, or (more minimally) a $\bZ_3$ symmetry taking $\{\ell,E,N\}\to \omega_3\{\ell,E,N\}$ and $\{\bar{e},\bar{E},\bar{n},\bar{N}\}\to \omega_3^{-1} \{\bar{e},\bar{E},\bar{n},\bar{N}\}$ where $\omega_k \equiv e^{2 \pi i/k}$. The resulting model is closely analogous to the quark sector described above with the $\bZ_3$ symmetry analogous to the $\bZ_3$ center of $\SU(3)_C$ (but without an analogue for $\bar{U}\bar{D}\bar{D}$). Due to this analogy, we omit further details.

\subsection{Gauged flavor symmetries}

There are two important features of the quark superpotential~(\ref{eq:quarkW}) which remain to be explained. Firstly, we must explain why the couplings $\lambda_{u,d}$ and $\lambda_{bnv}$ are flavor universal, as this is needed to obtain the MFV SUSY superpotential after integrating out the heavy fields. Moreover, we must also explain the absence of other flavor universal couplings, such as $\bar{u}\bar{d}\bar{d}$ and $\ell \ell \bar{e}$, which lead to unsuppressed baryon and/or lepton number violation. Phrased differently, we have both a ``flavor problem'' (explaining the flavor structure of certain couplings) and a problem of accidental symmetries (explaining the absence of certain couplings). These problems are related to but not synonymous with the usual problems of flavor and baryon/lepton number violation in the MSSM.

In this subsection, we focus on the first of these two problems, returning to the second issue later on. A crucial observation is that all the marginal couplings are flavor universal. This suggests the presence of a spontaneously broken flavor symmetry, where the nontrivial flavor structure of the mass terms descends from a marginal coupling to a flavor-Higgs superfield. Non-universal contributions to marginal couplings can still descend from non-renormalizable couplings to the flavor-Higgs field, but these are suppressed by $v_F/\Lambda$, where $v_F$ is the scale of flavor symmetry breaking and $\Lambda$ is the cutoff of the flavor-symmetric theory.

To avoid dangerous Goldstone modes (``familons") from the breaking of the flavor symmetry $G$ --- and also to protect $G$ from gravitational effects --- we choose to gauge it. We must therefore cancel the additional gauge anomalies $G^2 \U(1)_Y$ and $G^3$. While the former anomaly can be canceled by introducing additional ``exotic'' hypercharged matter, such fields are hard to remove from the low-energy spectrum and also hard to eventually embed into a GUT-like theory. We therefore wish to avoid introducing such exotic matter.   It is surprisingly easy to achieve this if only a diagonal subgroup is gauged. A further benefit of introducing the minimum amount of additional gauge symmetries is the ability to write down a relatively simple yet suitable rich Higgs potential for the flavor sector, as we explore in sections~\ref{sec:FCNC} and~\ref{sec:completemodel}. The simplest possibility is to gauge a diagonal $\SU(3)_Q$ for quark flavor  and a diagonal $\SU(3)_L$ for lepton flavor. Once this is achieved, it is easy to take a single diagonal anomaly-free $\SU(3)_F$ subgroup of the two to further simplify the model. 

Examining the marginal couplings in~(\ref{eq:quarkW}), we conclude that $q$, $\bar{U}$, and $\bar{D}$ transform under a common $\SU(3)_Q$ symmetry in the $\fund$, $\ov\fund$ and $\ov\fund$ representations, respectively. If we also require the couplings $\mu_{u,d}$ to be flavor universal, then we conclude that $\bar{u},\bar{d}$ and $U,D$ occupy conjugate representations, whereas $U,D$ and $\bar{U},\bar{D}$ must occupy \emph{the same} representation, otherwise $\cM_{u,d}$ would also be flavor universal. Applying the same considerations in the lepton sector leads to the charge table

\begin{center}
  \begin{tabular}{|c|ccc|cc|} \hline
    & $\SU(3)_C$ & $\SU(2)_L$ & $\U(1)_Y$ & $\SU(3)_Q$ & $\SU(3)_L$ \\ \hline 
    $q$ & $\fund$ & $\fund$ & $1 / 6$ & $\fund$ & $\sing$ \\
    $\bar{u}$ & $\ov{\fund}$ & $\sing$ & $- 2 / 3$ & $\fund$ & $\sing$ \\
    $\bar{d}$ & $\ov{\fund}$ & $\sing$ & $1 / 3$ & $\fund$ & $\sing$\\ \hline 
    $\ell$ & $\sing$ & $\fund$ & $- 1 / 2$ & $\sing$ & $\fund$ \\
    $\bar{e}$ & $\sing$ & $\sing$ & $1$ & $\sing$ & $\fund$ \\ \hline \hline
    $\bar{U}$ & $\ov{\fund}$ & $\sing$ & $- 2 / 3$ & $\ov\fund$ & $\sing$\\
    $U$ & $\fund$ & $\sing$ & $2 / 3$ & $\ov\fund$ & $\sing$\\
    $\bar{D}$ & $\ov{\fund}$ & $\sing$ & $1 / 3$ & $\ov\fund$ & $\sing$\\
    $D$ & $\fund$ & $\sing$ & $- 1 / 3$ & $\ov\fund$ & $\sing$\\ \hline 
    $\bar{E}$ & $\sing$ & $\sing$ & $1$ & $\sing$ & $\ov\fund$\\
    $E$ & $\sing$ & $\sing$ & $- 1$ & $\sing$ & $\ov\fund$\\
    $\bar{N}$ & $\sing$ & $\sing$ & $0$ & $\sing$ & $\ov\fund$\\ \hline
  \end{tabular}
\end{center}

Remarkably, all anomalies vanish, so there is no need to introduce exotic matter.


A variant of the lepton sector (also anomaly-free) with vector-like left-handed lepton doublets can be obtained by replacing the last three rows of the above table with

\begin{center}
  \begin{tabular}{|c|ccc|cc|} \hline
    & $\SU(3)_C$ & $\SU(2)_L$ & $\U(1)_Y$ & $\SU(3)_Q$ & $\SU(3)_L$ \\ \hline 
    $\bar{L}$ & $\sing$ & $\fund$ & $1/2$ & $\sing$ & $\ov\fund$\\
    $L$ & $\sing$ & $\fund$ & $- 1/2$ & $\sing$ & $\ov\fund$\\
    $\bar{n}$ & $\sing$ & $\sing$ & $0$ & $\sing$ & $\fund$\\ \hline
  \end{tabular}
\end{center}

A second variant of the lepton sector can be used if one wishes to obtain Dirac neutrino masses. In this case the lepton sector would contain the fields

\begin{center}
  \begin{tabular}{|c|cc|cc|} \hline
     & $\SU(2)_L$ & $\U(1)_Y$ & $\SU(3)_L$ & $\bZ_3$ \\ \hline 
  $\ell$ & $\fund$ & $-1/2$  &  $\fund$ & $\omega_3$ \\
  $\bar{e}$ & $\sing$ & $1$  &  $\fund$ & $\omega_3^{-1}$ \\
  $\bar{n}$ & $\sing$ & $0$  &  $\fund$ & $\omega_3^{-1}$ \\
  $\bar{E}$ & $\sing$ & $1$  &  $\ov\fund$ & $\omega_3^{-1}$ \\
  $E$ & $\sing$ & $-1$  &  $\ov\fund$ & $\omega_3$ \\
  $\bar{N}$ & $\sing$ & $0$  &  $\ov\fund$ & $\omega_3^{-1}$ \\
  $N$ & $\sing$ & $0$  &  $\ov\fund$ & $\omega_3$ \\
     \hline
  \end{tabular}
\end{center}

Here $\bZ_3$ is a subgroup of lepton number which can be gauged to forbid Majorana neutrino masses as well as the most dangerous lepton number violating operators. Note that all anomalies (including the discrete anomalies $\SU(2)^2 \bZ_3$, $\SU(3)_L^2 \bZ_3$ and $({\rm grav})^2 \bZ_3$) cancel. 

Having chosen one of these simple anomaly-free spectra, there are two different straightforward embeddings of $\SU(3)_F \subset \SU(3)_Q\times \SU(3)_L$: in one case all SM fields are $\SU(3)_F$ fundamentals (the ``standard embedding"), and in the other case the SM leptons are fundamentals while the quarks are anti-fundamentals (the ``flipped embedding"). The standard embedding, which we focus on, could potentially arise in a GUT-like theory, since all SM matter fields have the same flavor quantum numbers. However, we will not pursue complete GUT-like models in this paper, leaving this for future works~\cite{future}.

\subsection{The flavor Higgs sector and FCNCs\label{sec:FCNC}}

Given the matter content outlined above we still need to specify a flavor Higgs sector that is capable of completely breaking the flavor symmetry and producing the superpotential of (\ref{eq:quarkW},\ref{eq:leptonW}). To produce the large masses for the $U,\bar{U}$ and $D,\bar{D}$ heavy quarks we require flavor Higgs fields $\Phi_{u,d}$ in the $\rd{6}$ (symmetric) representation of the $\SU(3)_Q$ flavor symmetry. Since the anomalies of the matter fields all cancel, we assume that the flavor Higgs sector is vector-like, implying the existence of fields $\bar{\Phi}_{u,d}$ in the $\bar{\rd{6}}$ representation of $\SU(3)_Q$ as well. We likewise require Higgs fields in the $\rd{6}$ and $\bar{\rd{6}}$ representations of $\SU(3)_L$ to give masses to the heavy vector-like leptons and to generate a Majorana mass for the right-handed neutrinos. We label these fields as $\Phi_{e,\ell,n}$ or $\bar{\Phi}_{e,\ell,n}$ depending on whether they occupy a $\rd{6}$ or $\bar{\rd{6}}$ of $\SU(3)_L$ and on which SM fields they give a mass to. Finally, it is convenient (though not strictly necessary) to replace the parameters $\mu_{u,d,e,\ell}$ with singlet Higgs fields $\phi_{u,d,e,\ell}$. These fields will become charged fields when we later introduce discrete symmetries, and thus will also require vectorlike partners $\bar{\phi}_{u,d,e,\ell}$.

The flavor Higgs sector is then given by
\begin{equation}
  \begin{array}{|c|cc|} \hline
    & \SU(3)_Q & \SU(3)_L \\ \hline 
\Phi_{u,d} & \symm & \sing \\
\Phi_{e,n} & \sing & \symm   \\
 \phi_{u,d,e} & \sing & \sing \\ \hline
  \end{array}
\end{equation}
for the case with vector-like RH leptons $E,\bar{E}$, where we only show those Higgs fields required to give masses to the matter fields (and not their vector-like partners). The superpotential is now:
\begin{eqnarray} \label{eqn:WEEbar}
  W & = & \lambda_u q \bar{U} H_u + \lambda_d q \bar{D} H_d +
  \lambda_n \ell \bar{N} H_u + \lambda_e \ell \bar{E} H_d + \lambda_b \bar{U}
  \bar{D}  \bar{D} + \lambda_h S H_u H_d + \lambda_s S^3 \nonumber \\
  &  & + \Phi_u U \bar{U} + \Phi_d D \bar{D} + \Phi_e E \bar{E} + \Phi_n 
  \bar{N}^2 + \phi_u U \bar{u} + \phi_d D \bar{d} + \phi_e E \bar{e}
\end{eqnarray}
where we introduce one or more NMSSM singlet fields $S$. The case with vector-like lepton doublets $L,\bar{L}$ is quite similar, except that $\bar{\Phi}_n$ generates the neutrino Majorana mass rather than $\Phi_n$ due to the difference in $\SU(3)_L$ representations:
 \begin{eqnarray} \label{eqn:WLLbar}
  W  & = & \lambda_u q \bar{U} H_u + \lambda_d q \bar{D} H_d +
  \lambda_n L \bar{n} H_u + \lambda_e L \bar{e} H_d + \lambda_b  \bar{U} 
  \bar{D}  \bar{D} + \lambda_h S H_u H_d + \lambda_s S^3\nonumber \\
  &  & + \Phi_u U \bar{U} + \Phi_d D \bar{D} + \Phi_{\ell} L \bar{L} +
  \bar{\Phi}_n  \bar{n}^2 + \phi_u U \bar{u} + \phi_d D \bar{d} + \phi_{\ell} 
  \bar{L} \ell
\end{eqnarray}

We assume the presence of a Higgs potential which fixes all the moduli supersymmetrically and generates the required hierarchical Yukawa couplings. It is beyond the scope of this work to construct an explicit potential which does all of these things, but we can still impose minimum consistency requirements. To avoid pseudo-Goldstone bosons, we require a Higgs superpotential whose continuous symmetry group is precisely the (complexified) flavor gauge symmetry and no larger, and whose F-term conditions do not trivially set the vevs to zero. For instance, in the case of a single $\rd{6}\oplus\bar{\rd{6}}$ pair $\Phi,\bar{\Phi}$, the following potential meets all of these minimum requirements:
\begin{equation}
W = M \Phi \bar{\Phi} + \lambda \Phi^3 + \bar{\lambda} \bar{\Phi}^3 \,.
\end{equation}
Although one can show that this potential generates no hierarchies, it should be possible to generate hierarchies from the analogous but richer potential arising from multiple $\rd{6}\oplus\bar{\rd{6}}$ pairs. However, we will not attempt to do so explicitly in this work.

The absence of flavor-changing neutral currents (FCNCs) beyond those predicted by the SM sets a lower bound on the scale at which the $\SU(3)_F$ is Higgsed. In particular, the massive flavor gauge bosons generate the effective K\"ahler potential
\begin{equation}
K_{\rm eff} \sim g_F^2 [M^2]^{-1}_{a b} (q^{\dag} T^a q) (\bar{d}^{\dag} T^b \bar{d}) +\ldots
\end{equation}
where $T^a$ denotes an $\SU(3)_F$ generator, $g_F$ the flavor gauge coupling, and $M^2_{a b}$ the squared mass matrix for the flavor gauge bosons. Since we have only gauged a diagonal subgroup of the $\SU(3)^3$ MFV flavor symmetry, this operator contributes directly to $K$--$\bar{K}$ mixing even if $M^2_{a b}$ is $\SU(3)_F$ invariant. Thus, we can only suppress FCNCs by raising the flavor Higgsing scale $M/g_F \sim \langle \Phi \rangle$.

Specifically, generic constraints on CP violating $K$--$\bar{K}$ mixing require the new physics scale to exceed approximately $5\times10^5\mathrm{\ TeV}$ whereas generic constraints on CP conserving $K$--$\bar{K}$ mixing require the new physics scale to exceed approximately $3\times10^4\mathrm{\ TeV}$~\cite{bonatalk}. To avoid these constraints we conservatively require the flavor gauge bosons which interact with the down quark to be Higgsed at a scale $10^6 \mathrm{\ TeV}$ or higher, preventing excessive contributions to either $K$--$\bar{K}$ or $B$--$\bar{B}$ mixing. This can be accomplished by taking the greatest eigenvalue of $\langle \Phi^{m n}_d \rangle$ --- necessarily flavor-aligned with the down quark --- to be at least $10^6 \mathrm{\ TeV}$. While $B_s$--$\bar{B}_s$ and (due to CKM mixing) $D$--$\bar{D}$ mixings can be mediated by other flavor gauge bosons, the constraints on these processes are much weaker, requiring a new physics scale of at least $6\times10^2\mathrm{\ TeV}$ for $B_s$--$\bar{B}_s$ mixing and at least $6\times10^3\mathrm{\ TeV}$ for $D$--$\bar{D}$ mixing. The relatively small hierarchy between the down and strange quark masses ensures that the next largest eigenvalue of $\langle \Phi^{m n}_d \rangle$ be not less than $10^4 \mathrm{\ TeV}$, easily satisfying these constraints.

Alternately, we can accommodate a much smaller $\langle \Phi_d \rangle$ vev if $\SU(3)_F$ is completely broken at $10^6 \mathrm{\ TeV}$ or higher by anarchic neutrino masses $\langle\Phi_n\rangle$ or by another flavor-Higgs field. However, if $\langle\Phi_u\rangle$ is the dominant source of $\SU(3)_F$ breaking its largest eigenvalue must be substantially higher than this, due to the CKM misalignment between the up and down quarks. Because of this misalignment, certain dangerous flavor gauge bosons contributing to $K$--$\bar{K}$ mixing will only receive a mass at the scale of the second largest eigenvalue of $\langle\Phi_u\rangle$. Due to the large hierarchy between the charm quark and the up quark, this implies that the largest $\langle\Phi_u\rangle$ eigenvalue be at least $10^8 \mathrm{\ TeV}$ in this situation.

Due to this and to the large hierarchy between the up and top quarks, an LHC accessible up-type $\bar{u}^3$, $U_3$ vector-like pair is somewhat better motivated than the down-type equivalent in this scenario, though either can be achieved in certain limits.

In principle the massive flavor-Higgs fields $\Phi$, $\bar{\Phi}$ and $\phi$, $\bar{\phi}$ can also contribute to FCNCs as well as the flavor gauge bosons. However, since their interactions invariably involve vectorlike partners (such as $U$ and $D$) with negligible overlap with the light quarks, such contributions are at least loop suppressed, if not more. Furthermore, the masses of the uneaten Higgs fields are a priori unrelated to the Higgsing scale, and can in principle be made as heavy as necessary by choosing an appropriate Higgs potential. As such, we omit further discussion of this issue.

\section{Dangerous lepton and baryon-number violating operators} \label{sec:dangerousLNV}

The final missing component of our model is an explanation for the absence of dangerous superpotential terms which lead to excessive lepton number violation (LNV) or baryon number violation (BNV). For instance, in addition to the desired $\bar{U}\bar{D}\bar{D}$ superpotential operator, $\SU(3)_F$ flavor gauge invariance also allows the dangerous operators $\bar{u}\bar{d}\bar{d}$ and $\ell\ell\bar{e}$, which lead to unsuppressed BNV and LNV, respectively. Dangerous LNV can also be generated by higher-dimensional Planck-suppressed operators, such as $\frac{1}{\Lambda} \Phi \ell \ell \bar{E}$ or $\frac{1}{\Lambda} \Phi L \ell \bar{e}$, and both LNV and BNV can be generated upon integrating out the heavy flavors, such as via the operators $\bar{N} U \bar{U}$ and $U D D$.

Our approach is to introduce a discrete gauge symmetry, analogous to R-parity in the R-parity conserving MSSM, to forbid all problematic operators. Unlike its analogue, this discrete gauge symmetry is necessarily broken by the flavor Higgs fields, so there is no remnant in the low energy theory.

In this section, we aim to catalog the most dangerous operators in the high energy theory (both renormalizable and Planck-suppressed) which must be forbidden by this discrete symmetry. We do not attempt an exhaustive classification of all possible dangerous operators, since this list will depend on the flavor scale, superpartner masses, $\tan \beta$, and other details of the theory. Rather, we will list those operators which are obviously problematic, and which we will insist are forbidden by the discrete symmetry. Later, once we have chosen a discrete symmetry, we perform a more exhaustive search for LNV and BNV corrections.

\subsection{BNV operators}

We begin by discussing operators which violate baryon number only. The principle constraint on these operators is that they not induce too-rapid dinucleon decay.\footnote{As in~\cite{MFVSUSY}, bounds on $n$ -- $\bar{n}$ oscillation typically provide a subleading constraint.} For instance, if the low-energy effective BNV operator is $\bar{u}\bar{d}\bar{d}$, then applying the arguments of section 4.2 of~\cite{MFVSUSY} for a $\lambda''$ coupling with generic flavor structure, we see that if
\begin{equation}
\lambda''_{i j k} \lsim 10^{-8} \mathrm{\ \ for\ all\ } i,j,k\,,
\end{equation}
then dinucleon decay is sufficiently suppressed, where the exact bound depends somewhat on the superpartner masses and other details. While the bound actually applies to the $\lambda_{u d s}''$ coupling, other couplings will be less strongly constrained, as will higher-dimensional BNV effective operators.

Any Planck-suppressed operator in the high energy theory is necessarily suppressed by at least $\langle\Phi\rangle/M_{\rm pl} \sim 10^{-10}$ if we assume a flavor scale of $10^6\mathrm{\ TeV}$ in compliance with FCNC constraints, as discussed above. Thus, Planck-suppressed operators which violate only baryon number are not dangerous, whereas the only possible renormalizable BNV operators are
\begin{equation}
W_{\rm BNV} = \bar{U} \bar{D} \bar{D} + \bar{u}\bar{d}\bar{d}+U D D \,.
\end{equation}
The first of these operators leads to the MFV SUSY superpotential, as we have already shown, whereas the second leads to unsuppressed BNV in the low energy theory, and must be forbidden by the discrete symmetry. To determine the effect of the third operator, we must integrate out the heavy vector-like fields. Doing so in~(\ref{eq:quarkW}), we obtain:
\begin{equation} \label{eq:effectiveU}
U \to \frac{1}{\mu_u} (q H_u) \sqrt{1-\frac{Y_u Y_u^{\dag}}{|\lambda_u|^2}} Y_u + \frac{1}{2 \mu_u} w'' \left[\sqrt{1-\frac{Y_u Y_u^{\dag}}{|\lambda_u|^2}} Y_u\right] (Y_d \bar{d})^2 \,,
\end{equation}
and an analogous expression for $D$.
Thus, $UDD$ generates the effective operator\footnote{Strictly speaking, introducing $U D D$ will modify~(\ref{eq:effectiveU}), but these modifications only generate very high dimensional corrections and/or affect the numerical prefactors of the low energy effective operators, and can therefore be ignored.}
\begin{equation} \label{eq:effectiveUDD}
U D D \to \frac{1}{\mu_u \mu_d^2} \left[(q H_u) \sqrt{1-\frac{Y_u Y_u^{\dag}}{|\lambda_u|^2}} Y_u \right] \left[(q H_d) \sqrt{1-\frac{Y_d Y_d^{\dag}}{|\lambda_d|^2}} Y_d \right]^2 + \ldots
\end{equation}
where the omitted terms conserve baryon number and/or are subleading. Thus, we obtain a BNV operator with a pseudo MFV SUSY structure, though not strictly MFV.\footnote{Due to the presence of non-MFV terms in the superpotential, we must take the more general ansatz $Y_u = \mathrm{diag}(y_u,y_c,y_t) V_u$ and $Y_d = V_{CKM} \mathrm{diag}(y_d,y_s,y_b) V_d$, where $V_u$ and $V_d$ are in-principle arbitrary unitary matrices which can no longer be rotated away due to the reduced $\SU(3)_F \subset \SU(3)_q\times\SU(3)_{\bar{u}}\times\SU(3)_{\bar{d}}$ invariance; the combination $V_u V_d^{\dag}$ then appears in~(\ref{eq:effectiveUDD}).} Due to this structure and the $(v_u/\mu_u) (v_d/\mu_d)^2$ suppression, this operator should not induce excessive dinucleon decay.

Thus, of all possible BNV operators in the high energy theory, we find that only one operator need be forbidden:
\begin{equation} \label{eq:badBNV}
W_{\rm bad}^{(BNV)} = \bar{u}\bar{d}\bar{d} \,.
\end{equation}
While other non-MFV operators (if present) could still contribute to proton decay in the presence of lepton number violation, this is a model-dependent question which we defer until we present a complete model in section~\ref{sec:completemodel}.

\subsection{Low energy constraints on LNV operators} \label{subsec:LNVconstraints}

We now discuss operators which violate lepton number, including both baryon number conserving and violating variants. These operators can be generated in three possible ways. They can be either directly generated in the high energy theory, induced by vevs of the flavor Higgs fields, or generated upon integrating out the vectorlike flavors. In either of the first two cases, the resulting effective operators are either renormalizable or Planck suppressed, whereas the last mechanism will generate higher-dimensional operators with a lower cutoff. For the first two cases, it is expedient to classify all possible dangerous LNV corrections to the low energy effective theory that are either renormalizable or Planck suppressed and derive experimental bounds on these operators. These bounds can then be used to constrain the high-energy theory. We now present such a classification, returning to the question of LNV induced by integrating out the vectorlike flavors later.

Assuming that the right-handed neutrinos are heavy, and therefore absent from the low energy effective theory, we find the following potentially dangerous corrections to the MFV SUSY effective superpotential:\footnote{We omit the NMSSM singlet $S$ and the gauge invariant combination $H_u H_d$ in favor of their vevs, as this simplification will not affect the resulting bounds.}
\begin{equation}
  W_{\rm eff}^{\rm (LNV)} = \bar{\mu} \ell H_u + \lambda \ell \ell \bar{e} +\lambda'  q \ell
  \bar{d} + \frac{\tilde{\lambda}}{\Lambda} q^3 \ell + \frac{\tilde{\lambda}'}{\Lambda} q \bar{u} \bar{e} H_d +\frac{ \tilde{\lambda}''}{\Lambda} \bar{u} \bar{u} \bar{d} \bar{e}
\label{eq:effLNV}
\end{equation}
where dimension-six operators are sufficiently suppressed to avoid too-rapid proton decay.

We now discuss the experimental constraints on these couplings from the nonobservation of proton decay. We will assume that $\bar{u}\bar{d}\bar{d}$ has the MFV SUSY form~(\ref{eq:MFVSUSY}) to leading order along with MFV soft terms, whereas we take the lepton-number violating couplings to have a generic (non-MFV) flavor structure.

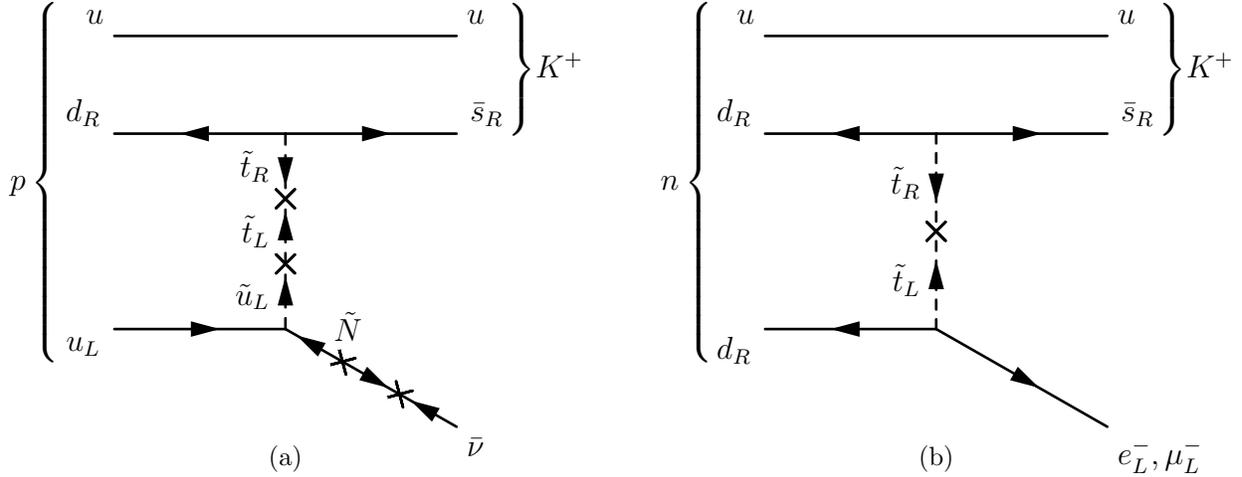
\begin{figure}
\begin{center}
\setlength{\unitlength}{0.65mm}
\subfigure[\label{fig:bilineardecay}]{
\begin{fmffile}{protondecay}
	        \begin{fmfgraph*}(70,80)
	     \fmfstraight
	        \fmfleft{i0,i1,i1_5,i2,i3}
	        \fmfright{o0,o1,o1_5,o2,o3}
	        \fmf{plain}{o3,i3}
	        \fmf{fermion}{v2,i2}
	        \fmf{fermion}{v2,o2}
	        \fmf{fermion}{i1,v1}
	        \fmf{phantom}{v1,o1}
	        \fmffreeze
	        \fmf{scalar,label=$\tilde{u}_L$,label.side=left}{v1,v4}
	        \fmf{scalar,label=$\tilde{t}_L$,label.side=left}{v4,v6}
	        \fmf{scalar,label=$\tilde{t}_R$,label.side=right}{v2,v6}
	        \fmf{fermion}{v5,v1}
	        \fmf{fermion}{v5,v7}
	        \fmf{fermion}{o0,v7}
	        \fmflabel{$u_L$}{i1}
	        \fmflabel{$d_R$}{i2}
	        \fmflabel{$u$}{i3}
	        \fmflabel{$u$}{o3}
	        \fmflabel{$\bar{s}_R$}{o2}
	        \fmflabel{$\bar{\nu}$}{o0}
	        \fmfv{decoration.shape=cross,decoration.angle=0,decoration.size=5thick}{v4}
	        \fmfv{decoration.shape=cross,decoration.angle=0,decoration.size=5thick}{v6}
	        \fmfv{decoration.shape=cross,decoration.angle=-30,decoration.size=5thick,label=$\tilde{N}$,label.angle=75,label.dist=8}{v5}
	        \fmfv{decoration.shape=cross,decoration.angle=-30,decoration.size=5thick}{v7}
\end{fmfgraph*}
\end{fmffile}
\rpar{K^+}{10mm}{46.5mm}{4mm}
\lpar{p}{25mm}{31.5mm}{-63.5mm}
}\hspace*{3cm}
\subfigure[\label{fig:qlddecay}]{
\begin{fmffile}{neutrondecay}
	        \begin{fmfgraph*}(70,80)
	     \fmfstraight
	        \fmfleft{i0,i1,i1_5,i2,i3}
	        \fmfright{o0,o1,o1_5,o2,o3}
	        \fmf{plain}{o3,i3}
	        \fmf{fermion}{v2,i2}
	        \fmf{fermion}{v2,o2}
	        \fmf{fermion}{v1,i1}
	        \fmf{phantom}{v1,o1}
	        \fmffreeze
	        \fmf{scalar,label=$\tilde{t}_L$,label.side=left}{v1,v4}
	        \fmf{scalar,label=$\tilde{t}_R$,label.side=right}{v2,v4}
	        \fmf{fermion}{v1,o0}
	        \fmflabel{$d_R$}{i1}
	        \fmflabel{$d_R$}{i2}
	        \fmflabel{$u$}{i3}
	        \fmflabel{$u$}{o3}
	        \fmflabel{$\bar{s}_R$}{o2}
	        \fmflabel{$e_L^-,\mu_L^-$}{o0}
	        \fmfv{decoration.shape=cross,decoration.angle=0,decoration.size=5thick}{v4}
\end{fmfgraph*}
\end{fmffile}
\rpar{K^+}{10mm}{46.5mm}{4mm}
\lpar{n}{25mm}{31.5mm}{-63.5mm}
}
\setlength{\unitlength}{1mm}
\end{center}
\caption{\subref{fig:bilineardecay} The leading contribution to proton decay  $p \to K^+ \bar{\nu}$ constraining the bilinear RPV term $\bar{\mu}$ from~\cite{MFVSUSY}. \subref{fig:qlddecay} The leading contribution to neutron decay yielding the strongest bound on the $\lambda'$ vertex.\label{fig:nucleondecay}}
\end{figure}

Bounds on bilinear LNV were discussed in detail in~\cite{MFVSUSY}, which in the present context gives\footnote{The bound given in~\cite{MFVSUSY} constrains the corresponding B-term, and consequently has a slightly different $\tan\beta$ dependence.}
\begin{equation}
w'' \bar{\mu} \lsim 4 \times 10^{-14} \frac{m_{\tilde{N}}}{\tan^3 \beta} \left( \frac{m_{\tilde{N}}}{100 \ {\rm GeV}} \right) \left( \frac{m_{\tilde{q}}}{100 \ {\rm GeV}} \right)^2
\end{equation}
from the process shown in Fig.~\ref{fig:bilineardecay}, where $w''$ is the MFV SUSY BNV parameter from~(\ref{eq:MFVSUSY}).

The leading nucleon decay diagram induced by $\lambda'$ is shown in Fig.~\ref{fig:qlddecay}. We estimate the width as:
\begin{equation}
\Gamma_{n \to K^+ \ell^-} \sim \frac{m_p}{8 \pi} \left(w'' \lambda'\, \frac{m_d m_s}{m_t^2} \left(\frac{\tilde{\Lambda}}{m_{\tilde{q}}}\right)^2 \tan^2 \beta\right)^2
\end{equation}
which leads to the bound:
\begin{equation}
w'' \lambda' \lsim 8\times 10^{-19} \frac{1}{\tan^2 \beta} \left(\frac{m_{\tilde{q}}}{100\mathrm{\ GeV}}\right)^2
\end{equation}
for $\tilde{\Lambda} \sim 250 \mathrm{\ GeV}$ using the $5.7\times 10^{31}\mathrm{\ yrs}$ experimental lower bound on the $n\to K^+\mu^-$ partial lifetime~\cite{Beringer:1900zz}. Similar considerations apply to the $\tilde{\lambda}'$ coupling upon inserting the $H_d$ vev, giving the bound:
\begin{equation}
w'' \tilde{\lambda}' \lsim 0.05\, \frac{1}{\tan\beta} \left(\frac{m_{\tilde{q}}}{100\mathrm{\ GeV}}\right)^2 \left(\frac{\Lambda}{10^{19}\mathrm{\ GeV}}\right)\,.
\end{equation}

\begin{figure}
\begin{center}
\setlength{\unitlength}{0.7mm}
\subfigure[\label{fig:nuNmixing}]{
\begin{fmffile}{nuNmixing}
	        \begin{fmfgraph*}(70,50)
	     \fmfstraight
	        \fmfleft{isp0,i1,isp1,i2}
	        \fmfright{osp0,o1,osp1,o2}
                  \fmf{fermion}{i1,v1}
                  \fmf{fermion,label=$\tau_R$,label.side=left}{v2,v1}
                  \fmf{fermion,label=$\tau_L$,label.side=right}{v2,v3}
                  \fmf{fermion}{o1,v3}
               \fmf{scalar,right,tension=0.4,label=$\tilde{\tau}_L$,label.side=right}{v3,v1}             
 	        \fmfv{decoration.shape=cross,decoration.angle=0,decoration.size=5thick}{v2}
 	        \fmflabel{$\nu$}{i1}
	        \fmflabel{$\tilde{N}$}{o1}
                  \fmf{fermion}{i2,v4}
                  \fmf{fermion,label=$\tau_L$,label.side=left}{v5,v4}
                  \fmf{fermion,label=$\tau_R$,label.side=right}{v5,v6}
                  \fmf{fermion}{o2,v6}
                  \fmf{scalar,right,tension=0.4,label=$\tilde{\tau}_R$,label.side=right}{v6,v4}             
 	        \fmfv{decoration.shape=cross,decoration.angle=0,decoration.size=5thick}{v5}
 	        \fmflabel{$\nu$}{i2}
	        \fmflabel{$\tilde{N}$}{o2}
\end{fmfgraph*}
\end{fmffile}
}
\hspace*{2.5cm}
\subfigure[\label{fig:4bodydecay}]{
\begin{fmffile}{neutrondecay2}
	        \begin{fmfgraph*}(70,80)
      	     \fmfstraight
	        \fmfleft{i5,i4,i3,i2,i1,i0}
	        \fmfright{o5,o4,o3,o2,o1,o0}
	        \fmf{plain}{o1,i1}
	        \fmf{fermion}{v2,i2}
                 \fmf{plain}{v2,v2p}
                 \fmf{fermion}{v2p,o2}
	        \fmf{fermion}{i4,v1}
                 \fmf{fermion,tension=2,label=$\tilde{C}$,label.side=left}{v5,v1}
                 \fmf{fermion,tension=2,label=$\tilde{C}$,label.side=right}{v5,v6} 
	        \fmf{phantom}{v6,o4}
	        \fmffreeze
	        \fmf{scalar,label=$\tilde{t}_L$,label.side=left}{v1,v4}
	        \fmf{scalar,label=$\tilde{t}_R$,label.side=right}{v2,v4}
                 \fmf{fermion}{o5,v6}
	        \fmf{scalar,label=$\tilde{l}_L$,label.side=left}{v6,v7}
                 \fmf{fermion}{o3,v7}
                 \fmf{fermion}{o4,v7}
	        \fmflabel{$d_R$}{i2}
	        \fmflabel{$d_L$}{i4}
	        \fmflabel{$u$}{i1}
	        \fmflabel{$u$}{o1}
	        \fmflabel{$\bar{s}_R$}{o2}
	        \fmflabel{$e_R^-,\mu_R^-$}{o3}
	        \fmflabel{$\bar{\nu}$}{o4}
	        \fmflabel{$\bar{\nu}$}{o5}
	        \fmfv{decoration.shape=cross,decoration.angle=0,decoration.size=5thick}{v4}
 	        \fmfv{decoration.shape=cross,decoration.angle=0,decoration.size=5thick}{v5}
\end{fmfgraph*}
\end{fmffile}
\rpar{K^+}{9mm}{39mm}{4mm}
\lpar{n}{22mm}{26mm}{-68mm}
}
\setlength{\unitlength}{1mm}
\end{center}
\caption{\subref{fig:nuNmixing} Loop diagrams contributing to $\nu -\tilde{N}$ mixing using the $\lambda$ vertex. Bounds will be obtained by including this mixing inside the diagram in Fig.~\ref{fig:bilineardecay}. \subref{fig:4bodydecay} The leading contribution to neutron decay using the $\lambda$ vertex if $\lambda_{i j j} = 0$ for all $i, j$ and the slepton masses are aligned with the lepton masses.\label{fig:lambdabound}}
\end{figure}
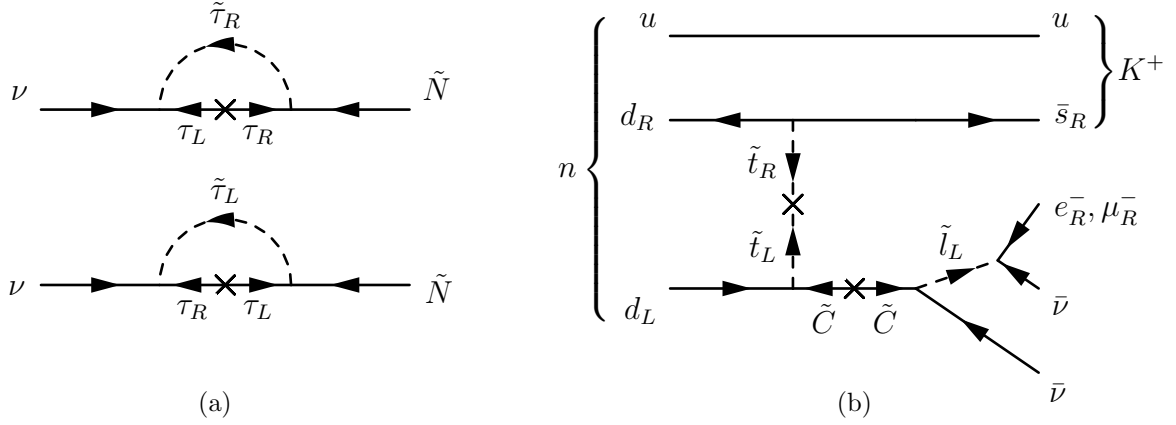

The leading contribution to nucleon decay induced by $\lambda$ comes from the loop diagram shown in Fig.~\ref{fig:nuNmixing}~\cite{Bhattacharyya:1998bx}, which gives a neutrino/neutralino mass mixing of order
\begin{equation}
\Delta m_{\nu \tilde{N}} \sim \frac{m_\tau}{16 \pi^2} \lambda \,.
\end{equation}
Applying the bilinear LNV constraints from~\cite{MFVSUSY}, we obtain the bound:
\begin{equation}
w'' \lambda \lsim 6 \times10^{-10} \frac{1}{\tan^4 \beta} \left( \frac{m_{\tilde{N}}}{100 \ {\rm GeV}} \right) \left( \frac{m_{\tilde{q}}}{100 \ {\rm GeV}} \right)^2
\end{equation}
However, the loop diagram vanishes if $\lambda_{i j j} = 0$ for all $i,j$, (e.g. for $\lambda_{i j k} \propto \epsilon_{i j k}$) if moreover the slepton masses are aligned with the charged lepton masses. In this case, the leading contribution to nucleon decay comes from the diagram shown in Fig.~\ref{fig:4bodydecay}. The width for the four-body decay is approximately
\begin{equation}
\Gamma_{n\to K^+ \ell^- \bar{\nu} \bar{\nu}} \sim \frac{2 \pi m_p^7}{(16 \pi^2)^3} \left(\lambda\, w''\frac{|V_{t d}| m_d m_s}{m_t^2} \left(\frac{\tilde{\Lambda}}{m_{\tilde{q}}}\right)^2 \frac{\tan^2 \beta}{m_{\tilde{N}} m_{\tilde{\ell}}^2}\right)^2 \,.
\end{equation}
While there is no direct bound on this decay mode, we assume a baseline sensitivity of at least $10^{30} \mathrm{\ yrs}$ (which is similar to the bound on neutron disappearance~\cite{Araki:2005jt}.) We then obtain the bound:
\begin{equation}
w'' \lambda \lsim 1.3 \times10^{-7} \frac{1}{\tan^2 \beta} \left( \frac{m_{\tilde{N}}}{100 \ {\rm GeV}} \right) \left( \frac{m_{\tilde{q}}}{100 \ {\rm GeV}} \right)^2 \left( \frac{m_{\tilde{\ell}}}{100 \ {\rm GeV}} \right)^2 \,,
\end{equation}
for this special case.

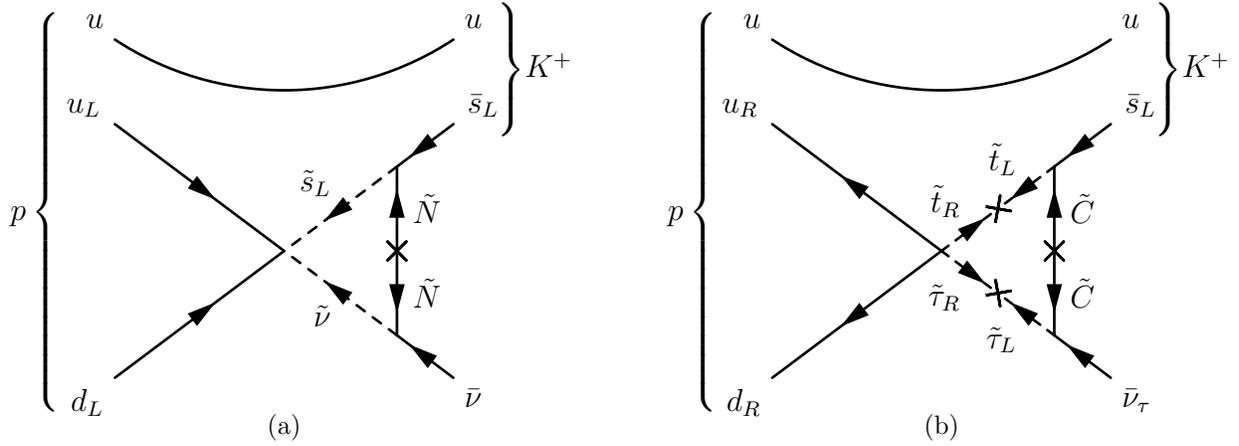
\begin{figure}
\begin{center}
\setlength{\unitlength}{0.5mm}
\subfigure[\label{fig:qqqldecay}]{
\begin{fmffile}{protondecay2}
	        \begin{fmfgraph*}(90,90)
      	     \fmfstraight
	        \fmfleft{i1,i4,i5,i2,i3}
	        \fmfright{o1,o4,o5,o2,o3}
	        \fmf{plain,left=0.3}{o3,i3}
	        \fmf{fermion}{i2,v1}
                 \fmf{fermion}{i1,v1}
                 \fmf{scalar,tension=1.5,label=$\tilde{s}_L$,label.side=right}{v2,v1}
                 \fmf{scalar,tension=1.5,label=$\tilde{\nu}$,label.side=left}{v4,v1}
                 \fmf{fermion,tension=0.01,label=$\tilde{N}$,label.side=right}{v3,v2}
                 \fmf{fermion,tension=0.01,label=$\tilde{N}$,label.side=left}{v3,v4}
                 \fmf{fermion,tension=3}{o2,v2}
                 \fmf{fermion,tension=3}{o1,v4}
 	        \fmfv{decoration.shape=cross,decoration.angle=0,decoration.size=5thick}{v3}
	        \fmflabel{$u_L$}{i2}
	        \fmflabel{$d_L$}{i1}
	        \fmflabel{$u$}{i3}
	        \fmflabel{$u$}{o3}
	        \fmflabel{$\bar{s}_L$}{o2}
	        \fmflabel{$\bar{\nu}$}{o1}
\end{fmfgraph*}
\end{fmffile}
\rpar{K^+}{9.5mm}{40mm}{3mm}
\lpar{p}{28mm}{21mm}{-63mm}
}
\hspace*{3cm}
\subfigure[\label{fig:uudedecay}]{
\begin{fmffile}{protondecay3}
	        \begin{fmfgraph*}(90,90)
      	     \fmfstraight
	        \fmfleft{i1,i4,i5,i2,i3}
	        \fmfright{o1,o4,o5,o2,o3}
	        \fmf{plain,left=0.3}{o3,i3}
	        \fmf{fermion}{v1,i2}
                 \fmf{fermion}{v1,i1}
                 \fmf{scalar,tension=3,label=$\tilde{\tau}_R$,label.side=right}{v1,v7}
                 \fmf{scalar,tension=3,label=$\tilde{\tau}_L$,label.side=left}{v4,v7}
                 \fmf{scalar,tension=3,label=$\tilde{t}_R$,label.side=left}{v1,v5}
                 \fmf{scalar,tension=3,label=$\tilde{t}_L$,label.side=right}{v2,v5}
                 \fmf{fermion,tension=0.01,label=$\tilde{C}$,label.side=right}{v6,v2}
                 \fmf{fermion,tension=0.01,label=$\tilde{C}$,label.side=left}{v6,v4}
                 \fmf{fermion,tension=3}{o2,v2}
                 \fmf{fermion,tension=3}{o1,v4}
 	        \fmfv{decoration.shape=cross,decoration.angle=35,decoration.size=5thick}{v5}
                 \fmfv{decoration.shape=cross,decoration.angle=0,decoration.size=5thick}{v6}
                 \fmfv{decoration.shape=cross,decoration.angle=-35,decoration.size=5thick}{v7}
	        \fmflabel{$u_R$}{i2}
	        \fmflabel{$d_R$}{i1}
	        \fmflabel{$u$}{i3}
	        \fmflabel{$u$}{o3}
	        \fmflabel{$\bar{s}_L$}{o2}
	        \fmflabel{$\bar{\nu}_{\tau}$}{o1}
\end{fmfgraph*}
\end{fmffile}
\rpar{K^+}{9.5mm}{40mm}{3mm}
\lpar{p}{28mm}{21mm}{-63mm}
}
\setlength{\unitlength}{1mm}
\end{center}
\caption{The leading contributions to proton decay from the higher dimensional R-parity even couplings \subref{fig:qqqldecay} $\tilde{\lambda}$ and \subref{fig:uudedecay} $\tilde{\lambda}''$.\label{fig:lambdatildebound}}
\end{figure}

The R-parity even couplings $\tilde{\lambda}$ and $\tilde{\lambda}''$ lead directly to proton decay independent of the BNV $w''$ coupling. For $\tilde{\lambda}$ the dominant diagram is shown in Fig.~\ref{fig:qqqldecay}, with the width
\begin{equation}
\Gamma_{p \to K^+ \bar{\nu}} \sim \frac{m_p}{8 \pi} \left(\frac{\tilde{\lambda} \tilde{\Lambda}^2}{16 \pi^2 \Lambda\, m_{\rm soft}}\right)^2 \,,
\end{equation}
which gives the bound
\begin{equation}
\tilde{\lambda} \lsim 4 \times 10^{-8} \left(\frac{m_{\rm soft}}{100\mathrm{\ GeV}}\right) \left(\frac{\Lambda}{10^{19}\mathrm{\ GeV}}\right) \,.
\end{equation}
For $\tilde{\lambda}''$ there is more flavor suppression (see Fig.~\ref{fig:uudedecay}), and we obtain the weaker bound:
\begin{equation}
\tilde{\lambda}'' \lsim 10^{-4} \frac{1}{\tan \beta} \left(\frac{m_{\rm soft}}{100\mathrm{\ GeV}}\right) \left(\frac{\Lambda}{10^{19}\mathrm{\ GeV}}\right) \,.
\end{equation}

\subsection{Directly induced lepton number violation}

Based on the above constraints on corrections to the low energy theory, we now search for LNV operators in the high energy theory which can violate these constraints. In this subsection, we focus on operators which directly induce lepton number violation in the low energy theory, deferring consideration of LNV operators containing the heavy fields $U,D,E,\bar{N}$ or $\bar{n},\bar{L}$ to the next section. 

To select operators which are potentially relevant, we consider the reference point $\tan \beta = 10$, $m_{\rm soft} = 300 \mathrm{\ GeV}$ and $w'' \sim 1$, with $\langle \Phi \rangle, \langle \bar{\Phi} \rangle\sim 10^6\mathrm{\ TeV}$ and $\langle \phi \rangle \lsim 10^3\mathrm{\ TeV}$ in accordance with the Yukawa hierarchies. We then consider all possible gauge invariant operators which can generate the operators in~(\ref{eq:effLNV}) upon inserting the flavor Higgs vevs, accounting for the flavor structure induced by the mass mixings and retaining all operators which violate the experimental constraints for an order one coefficient and a cutoff of $10^{19}\mathrm{\ GeV}$. Since $\langle \phi \rangle/\Lambda \sim 10^{-10}$, dimension six operators are sufficiently suppressed except in the case of the $\ell H_u$ coupling, and we can otherwise restrict our attention to dimension four and five operators.

The resulting list of dangerous gauge-invariant operators will depend on whether we choose the standard or flipped embedding of $\SU(3)_Q\times\SU(3)_L$ into $\SU(3)_F$. We find that the following dangerous operators are common to the two cases:
\begin{equation} \label{eq:badUnivDir}
W_{\rm bad} = \ell \ell \bar{e}+\frac{1}{\Lambda} \Phi \ell \ell \bar{E} + \frac{1}{\Lambda} \Phi L \ell \bar{e} + \frac{1}{\Lambda} \Phi q \ell \bar{D} + \frac{1}{\Lambda} \bar{\Phi} q L \bar{D} + \frac{1}{\Lambda^2} \Phi^2 \bar{\Phi} \ell H_u \,.
\end{equation}
With the standard embedding we have the additional dangerous operators
\begin{equation} \label{eq:badStdDir}
W_{\rm bad}^{\rm (standard)} = \left(1+\frac{1}{\Lambda} \phi + \frac{1}{\Lambda} S\right) q \ell \bar{d} +\frac{1}{\Lambda} \Phi q L \bar{d}+\frac{1}{\Lambda} q \bar{u} \bar{e} H_d+\frac{1}{\Lambda} \bar{u} \bar{u} \bar{D} \bar{E}+\frac{1}{\Lambda} \bar{U} \bar{u} \bar{d} \bar{E}+\frac{1}{\Lambda} \bar{u} \bar{U} \bar{D} \bar{e} \,,
\end{equation}
whereas with the flipped embedding, we have the additional dangerous operators
\begin{equation} \label{eq:badFlipDir}
W_{\rm bad}^{\rm (flipped)} = \left(1+\frac{1}{\Lambda} \phi \right) q L \bar{d} +\frac{1}{\Lambda} \bar{\Phi} q \ell \bar{d}+\frac{1}{\Lambda} \bar{u} \bar{u} \bar{D} \bar{e}+\frac{1}{\Lambda} \bar{U} \bar{u} \bar{d} \bar{e}+\frac{1}{\Lambda} \bar{u} \bar{U} \bar{D} \bar{E} \,.
\end{equation}
In each case, only some of these operators exist in a given theory, depending on which type of vector-like leptons are present.

These lists should be treated as representative only, since some operators on the list barely make the cut, such as $\frac{1}{\Lambda} \Phi \ell \ell \bar{E}$, and others barely miss it, such as $\frac{1}{\Lambda} \bar{U}\bar{U}\bar{d}\bar{e}$. Nonetheless, we will find that it is possible to forbid all of these operators (and many more besides) by choosing an appropriate discrete symmetry.

\subsection{Lepton number violation mediated by heavy flavors}

We now turn to the question of lepton number violation mediated by the heavy flavors, arising from LNV operators involving $U,D,E,\bar{N}$ or $U,D,\bar{L},\bar{n}$ (depending on the theory). We have already considered a BNV operator of this type in~(\ref{eq:effectiveUDD}), and we take the same approach in what follows, integrating out the heavy fields using the replacement~(\ref{eq:effectiveU}), its analogue for $D$, and the replacements
\begin{eqnarray} \label{eq:effectiveEN}
E & \to & \frac{1}{\mu_e} (\ell H_d) \sqrt{1-\frac{Y_e Y_e^{\dag}}{|\lambda_e|^2}} Y_e\nonumber \\
\bar{N} & \to & \frac{1}{\lambda_n} \frac{m_{\nu}}{v_u^2} (\ell H_u)
\end{eqnarray}
or
\begin{eqnarray} \label{eq:effectiveLn}
\bar{L} & \to & \frac{1}{\mu_{\ell}} \sqrt{1-\frac{Y_e Y_e^{\dag}}{|\lambda_e|^2}} \left[Y_e (\bar{e} H_d) + \frac{m_{\nu}}{v_u^2} (\ell H_u) H_u\right]\nonumber \\
\bar{n} & \to & \frac{\lambda_e}{\lambda_n}\, Y_e^{-1} \frac{m_{\nu}}{v_u^2} (\ell H_u)
\end{eqnarray}
depending on which type of vector-like leptons are present, where $m_{\nu}$ is the left-handed neutrino Majorana mass matrix generated by the effective operator $\frac{1}{v_u^2} (\ell H_u) m_{\nu} (\ell H_u)$.

Thus, to find dangerous operators, in principle all we need do is to list all LNV dimension four and five operators in the high energy theory that we have not  considered yet (those involving $U,D,E,\bar{L},\bar{N}$ or $\bar{n}$), making the above replacements and then considering the consequences of the resulting effective operator for the low energy theory. This list contains a much wider variety of effective operators than those considered above, and it is very lengthy to derive explicit bounds for every possible operator. Instead, we develop a heuristic scheme to estimate which operators are likely dangerous.

Except in a few special cases where the high-energy operator is super-renormalizable after inserting the flavor Higgs vevs, the strongest bounds will come from inserting the electroweak Higgs vevs into the replacements~(\ref{eq:effectiveU}, \ref{eq:effectiveEN}, \ref{eq:effectiveLn}), as this results in a lower-dimensional effective operator. Upon doing so for $U,D,E$ or $\bar{L}$ insertions, we obtain one of the light lepton or quark superfields suppressed by a factor of the mass of the corresponding fermion divided by $\mu_u$, $\mu_d$, $\mu_e$ or $\mu_{\ell}$ respectively, with a possible additional suppression for the third generation coming from the $\sqrt{1 - \frac{Y_x Y_x^{\dag}}{|\lambda_x|^2}}$ factor. In what follows, we assume that $\mu_x\gsim1\mathrm{\ TeV}$ for $x=u,d,\ell,e$, consistent with $\langle \Phi \rangle \sim 10^6 \mathrm{\ TeV}$ and the known Yukawa hierarchies.

For $U$ and $D$ there is an additional BNV term which can directly induce proton decay when inserted into a baryon-number conserving LNV operator. The resulting operator will be dimension five or higher, requiring a gaugino exchange loop to induce proton decay. This can be compared to the similar tree-level diagram involving squark exchange between the $\bar{u}\bar{d}\bar{d}$ MFV SUSY superpotential operator and the baryon-number conserving LNV operator. In place of the $m_q/\mu_{u,d}$ suppression from integrating out $U$ or $D$, the loop diagram has a $\frac{g^2 m_{\rm soft}}{16 \pi^2 \mu_{u,d}}$ suppression, but otherwise a very similar structure. For $m_{\rm soft} \sim 300\mathrm{\ GeV}$ the loop diagram only dominates in place of the exchange of a ``light'' ($u,d,s$) squark. Since such diagrams are typically suppressed for other reasons, the loop diagram is usually subdominant.

Now consider $\bar{N}$ insertions. If we assume $m_{\nu} \sim 0.1\mathrm{\ eV}$ then every such operator comes with a strong $m_{\nu}/v_u \sim 6\times 10^{-13}$ suppression. However, since we assume $\langle \Phi \rangle \sim 10^6\mathrm{\ TeV}$, we require $\mathcal{M}_n \lsim 10^{6}\mathrm{\ TeV}$ which implies that $\lambda_n \lsim 2\times 10^{-3}$. Taking this into account, we find an overall suppression factor of about $3\times 10^{-10}$ for each $\bar{N}$ insertion. A similar argument applies to $\bar{n}$, except that the minimum suppression per $\bar{n}$ insertion is now only about $10^{-7}$ due to the factor of $Y_e^{-1}$.

We now proceed to classify all possible operators of dimension five or less based on the number of leptons and quarks they contain. We need not consider operators which violate lepton number by an even number, as this will not induce proton decay, so we can have either one or three leptons. Operators with three leptons cannot have any quarks due to the restriction on dimensionality, whereas operators with one lepton can have zero, two, or three quarks, where the latter also violate baryon number. In general operators in the high energy theory and the resulting effective operators in the low energy theory will have the same number of quarks and leptons, except that operators with two quarks and a lepton can also generate operators with three quarks and a lepton in the low energy theory through the insertion of the second term in~(\ref{eq:effectiveU}).

We begin by considering operators with three leptons. Following the discussion in \S\ref{subsec:LNVconstraints}, we anticipate that a coupling of less than about $10^{-12}$ (roughly the bound on $\lambda$ at our chosen reference point) is sufficient to suppress  operators of this type to acceptable levels. Using this estimate, we find that none of the possible gauge invariant operators of this type (such as $\bar{N}^3, \bar{N} E \bar{E}$, etc.) are dangerous.

Next, we consider operators with one lepton and no quarks. It is straightforward to check that the only dangerous operators of this type are the RH neutrino tadpoles
\begin{equation} \label{eq:badUnivL}
W^{(L)}_{\rm bad} = \frac{1}{\Lambda} \Phi \bar{\Phi}^2 \bar{N}+\frac{1}{\Lambda} \Phi^2 \bar{\Phi} \bar{n}+\frac{1}{\Lambda^2} \Phi^4 \bar{n}+\frac{1}{\Lambda^2} \Phi \bar{\Phi}^3 \bar{n}
\end{equation}
where as usual only some of these operators will appear in a given theory, depending on whether $\bar{N}$ or $\bar{n}$ is present. These tadpoles, which induce bilinear lepton number violation in the low energy theory, are a special case where dimension-six operators, such as $\frac{1}{\Lambda^2} \Phi^4 \bar{n}$ can be (at least marginally) dangerous. Note that this operator differs from the analogous operator $\frac{1}{\Lambda^2} \bar{\Phi}^4 \bar{N}$, which is small enough by about a factor of 10 to avoid experimental constraints; the difference lies in the different right-handed neutrino Yukawa couplings implied by the two models. In any case, the dimension-six contribution to the tadpole may be made sufficiently small by lowering the flavor scale to $5\times 10^5\mathrm{\ TeV}$ (still in reasonable agreement with flavor constraints), so it is in fact not very dangerous.

Next, we consider operators with one lepton and two quarks. Based on the discussion in \S\ref{subsec:LNVconstraints}, we anticipate that a coupling of less than about $10^{-20}$ (roughly an order of magnitude smaller than the bound on $\lambda'$ at our chosen reference point, accounting for the possibility of the more strongly constrained $p \to K^+ \bar{\nu}$ decay mode) is sufficient to suppress operators of this type to acceptable levels. The dangerous gauge-invariant operators will depend on whether we choose the standard or flipped embedding. The following dangerous operators are common to the two cases:
\begin{equation} \label{eq:badUnivQQL}
W_{\rm bad} = \frac{1}{\Lambda} \Phi \bar{n} U \bar{u} +\frac{1}{\Lambda} \bar{\Phi} q \bar{U} \bar{L}+\frac{1}{\Lambda} \bar{\Phi}U\bar{d} E + \frac{1}{\Lambda} \bar{\Phi} \bar{u} D \bar{E}+ \frac{1}{\Lambda} \Phi \bar{u} D \bar{e}
\end{equation}
In the first case, we obtain the additional dangerous operators:
\begin{equation} \label{eq:badStdQQL}
W_{\rm bad}^{\rm (standard)} = \bar{N} U \bar{U}+\bar{N} D \bar{D} + U \bar{D} E + \bar{U} D \bar{E} + \frac{1}{\Lambda} \bar{\Phi} \bar{U} D \bar{e}+\frac{1}{\Lambda} \Phi q \bar{u} \bar{L}
\end{equation}
whereas in the flipped case, we obtain the additional dangerous operators:
\begin{equation} \label{eq:badFlipQQL}
W_{\rm bad}^{\rm (flipped)} = \bar{n} U \bar{U}+\bar{n} D \bar{D} +\left(1+\frac{1}{\Lambda} \phi\right)q \bar{u} \bar{L} + \left(1+\frac{1}{\Lambda} \phi\right) \bar{U} D \bar{e} +\frac{1}{\Lambda} \Phi U\bar{D} E + \frac{1}{\Lambda} \Phi \bar{U} D \bar{E}
\end{equation}

Finally, we consider operators with one lepton and three quarks, which are necessarily dimension five and require a loop to induce proton decay. Based on the discussion in \S\ref{subsec:LNVconstraints}, we expect that a coupling of less than $10^{-7}$ for a Planck scale cutoff (roughly the bound on $\tilde{\lambda}$ for our chosen parameters) is just sufficient to suppress proton decay to an acceptable level. Using this estimate, we obtain the following dangerous operators for the standard embedding
\begin{equation} \label{eq:badStdQQQL}
W_{\rm bad}^{\rm (standard)} = \frac{1}{\Lambda} q^2 U E+\frac{1}{\Lambda} \bar{d}^2 \bar{D} E \,,
\end{equation}
and none in the flipped embedding.

\section{A complete model using a discrete symmetry\label{sec:completemodel}}

Having enumerated the operators that are most likely to lead to proton decay or $\Delta B = 2$ processes (see~(\ref{eq:badBNV}), (\ref{eq:badUnivDir})--(\ref{eq:badFlipDir}), (\ref{eq:badUnivL})--(\ref{eq:badStdQQQL})) we now search for a discrete symmetry which forbids these operators. In addition to these dangerous LNV and BNV corrections, we also aim to prevent the problematic cross-couplings between the electroweak and flavor Higgs sectors:
\begin{equation} \label{eq:badEW1}
W_{\rm bad}^{(\rm cross)} = \mu_{\phi} \phi S+\phi H_u H_d+\phi S^2+\phi^2 S+\Phi \bar{\Phi} S + \frac{1}{\Lambda} \Phi^3 S+ \frac{1}{\Lambda} \bar{\Phi}^3 S \,,
\end{equation}
which can lead to large dimensionful couplings in the Higgs potential and hence fine tunings. We can also solve the usual $\mu$ problem by forbidding the super-renormalizable operators
\begin{equation} \label{eq:badEW2}
W_{\rm bad}^{(EW)} =  \hat{\mu}^2 S+ \mu H_u H_d + \mu_s S^2 \,.
\end{equation}

On the other hand, the discrete symmetry will also constrain the flavor Higgs potential, potentially leading to accidental symmetries in the flavor Higgs sector. Such accidental symmetries will induce dangerous Goldstone modes which could mediate FCNCs. Remarkably, we will show that it is possible to choose a discrete symmetry which satisfies all of these constraints while allowing for a semi-realistic flavor Higgs potential without accidental symmetries.

As this discrete symmetry is meant to constrain Planck suppressed as well as renormalizable couplings, it must be anomaly-free and gauged.\footnote{Discrete gauge symmetries sometimes appear as remnants of a spontaneously broken continuous gauge symmetry, but they are well defined and distinct from discrete global symmetries even in the absence of such a mechanism.} The discrete symmetry could be an ordinary symmetry or an R-symmetry. In the case of a discrete R-symmetry the superspace coordinate obtains a non-trivial phase $\eta_{\theta}$ under the discrete transformation, implying that gauginos are rotated by $\eta_{\theta}$ as well whereas the superpotential must pick up a phase $\eta_W= \eta_{\theta}^2$. In Appendix~\ref{sec:appendix} we show that the anomaly cancellation conditions for the discrete symmetry together with the requirement that the operators in~(\ref{eq:badEW1}) are forbidden requires a discrete R-symmetry. Focusing on the case with $E,\bar{E}$ leptons and the regular embedding, we further argue that the smallest order choice for a discrete symmetry group forbidding all problematic operators while allowing for a semirealistic flavor Higgs potential is a $\bZ_{11}$ discrete R-symmetry, where we assume that the flavor Higgs sector is completely vector-like.

We now present an example of a complete model with a discrete $\bZ_{11}$ R-symmetry. We choose $\eta_\theta = \omega_{11}^{-1} = e^{-2\pi i/11}$ without loss of generality, and thus $\eta_W= \omega_{11}^{-2}$. We then introduce the ``matter'' fields:
\begin{center}
 \begin{tabular}{|c|ccc|cc|}
    \hline
    & $\mathrm{SU} ( 3)_C$ & $\mathrm{SU} ( 2)_L$ & $U ( 1)_Y$ & $\mathrm{SU} (
    3)_F$ & $\mathbbm{Z}_{11}  [ \omega_{11}^{- 2}]$ \\ \hline
    $q$ & $\fund$ & $\fund$ & $1 / 6$ & $\fund$ & $\omega_{11}^3$\\
    $\bar{u}$ & $\ov{\fund}$ & $\sing$ & $- 2 / 3$ & $\fund$ & $\omega_{11}^4$\\
    $\bar{d}$ & $\ov{\fund}$ & $\sing$ & $1 / 3$ & $\fund$ & $\omega_{11}^5$\\
    $\ell$ & $\sing$ & $\fund$ & $- 1 / 2$ & $\fund$ & $\omega_{11}^4$\\
    $\bar{e}$ & $\sing$ & $\sing$ & $1$ & $\fund$ & $1$\\
    $\bar{U}$ & $\ov{\fund}$ & $\sing$ & $- 2 / 3$ & $\ov\fund$ & $\omega_{11}^3$\\
    $\bar{D}$ & $\ov{\fund}$ & $\sing$ & $1 / 3$ & $\ov\fund$ & $\omega_{11}^3$\\
    $\bar{E}$ & $\sing$ & $\sing$ & $1$ & $\ov\fund$ & $\omega_{11}^2$\\
    $U$ & $\fund$ & $\sing$ & $2 / 3$ & $\ov\fund$ & $\omega_{11}$\\
    $D$ & $\fund$ & $\sing$ & $- 1 / 3$ & $\ov\fund$ & $\omega_{11}^5$\\
    $E$ & $\sing$ & $\sing$ & $- 1$ & $\ov\fund$ & $\omega_{11}^4$\\
    $\bar{N}$ & $\sing$ & $\sing$ & $0$ & $\ov\fund$ & $\omega_{11}^2$ \\ \hline
    $H_u$ & $\sing$ & $\fund$ & $1 / 2$ & $1$ & $\omega_{11}^3$\\
    $H_d$ & $\sing$ & $\fund$ & $- 1 / 2$ & $1$ & $\omega_{11}^3$\\
    $S$ & $\sing$ & $\sing$ & $0$ & $1$ & $\omega_{11}^3$ \\ \hline
  \end{tabular}
\end{center}
and the flavor Higgs fields
\begin{center}
 \begin{tabular}{|c|cc|}
    \hline
    & $\mathrm{SU}(3)_F$ & $\mathbbm{Z}_{11}  [ \omega_{11}^{- 2}]$ \\ \hline
    $\Phi_{u,n}$   & $\symm$ & $\omega_{11}^5$\\
    $\Phi_d$   & $\symm$ & $\omega_{11}$\\
    $\Phi_e$   & $\symm$ & $\omega_{11}^3$\\
    $\bar{\Phi}_{u,n}$   & $\ov\symm$ & $\omega_{11}^4$\\
    $\bar{\Phi}_d$  & $\ov\symm$ & $\omega_{11}^8$\\
    $\bar{\Phi}_e$ & $\ov\symm$ & $\omega_{11}^6$\\
    $\phi_u$   & $1$ & $\omega_{11}^4$\\
    $\phi_d$  & $1$ & $\omega_{11}^{- 1}$\\
    $\phi_e$   & $1$ & $\omega_{11}^5$ \\ \hline
  \end{tabular}
\end{center}
where $\bZ_k [ \eta_W ] $ denotes a $\bZ_k$ discrete symmetry under which the superpotential picks up a phase $\eta_W$.\footnote{In specifying a discrete R-symmetry, it is unnecessary to specify $\eta_{\theta}$ if $\eta_W = \eta_{\theta}^2$ is given, since the two possible sign choices in taking the square root are related by $(-1)^F$.}

As shown in appendix~\ref{sec:appendix}, this model is anomaly free.\footnote{There is a naive $(\mathrm{grav})^2 \bZ_{11}$ anomaly which can be cancelled by adding a second copy of the $S$ field, but any hidden (e.g.\ SUSY-breaking) sector will contribute to this anomaly, as will the gravitino, so there is no clear constraint on the number of NMSSM singlets.}
The most general renormalizable flavor Higgs superpotential allowed by the $\bZ_{11}$ R-symmetry is:
\begin{eqnarray}
  W_{\mathrm{Higgs}} & = & M_u \Phi_u  \bar{\Phi}_u + M_d \Phi_d  \bar{\Phi}_d +
  M_e \Phi_e  \bar{\Phi}_e + \lambda_1 \phi_u \Phi_d  \bar{\Phi}_u + m_1
  \phi_u \phi_e + m_2 \phi_d^2 + \lambda_2 \phi_e^2 \phi_d \nonumber \\
  &  & + \lambda_{u d e} \Phi_u \Phi_d \Phi_e + \lambda_{e e e} \Phi_e^3 +
  \bar{\lambda}_{u d d}  \bar{\Phi}_u  \bar{\Phi}_d^2 + \bar{\lambda}_{d e e} 
  \bar{\Phi}_d  \bar{\Phi}_e^2 \label{eq:HiggsPotential}
\end{eqnarray}
where $\Phi_u$ now stands for either $\Phi_u$ or $\Phi_n$ (which carry the same charges). One can check that this potential breaks all $\U(1)$ global symmetries, and hence does not obviously lead to Goldstone modes. Although we may in general require more than one ``flavor'' of each type of $\Phi$ field to allow a suitably rich potential which can reproduce the flavor structure of the SM, the potential is likely sufficiently generic to also break any resulting nonabelian flavor symmetry, avoiding Goldstones. However, we will not study the flavor Higgs sector in detail, deferring this to a future work.

One can show that there are no further renormalizable superpotential couplings allowed by the $\bZ_{11}$ R-symmetry beyond those in (\ref{eqn:WEEbar}), (\ref{eq:HiggsPotential}). Performing a systematic search we find the following dimension-five lepton-number violating operators:
\begin{equation}
W_{\rm LNV}^{(5)} = \frac{1}{\Lambda} \bar{\Phi}_d \bar{N} D \bar{d} + \frac{1}{\Lambda} \phi_d \bar{N} D \bar{D}+\frac{1}{\Lambda} \phi_d \bar{U} D \bar{E}+ \frac{1}{\Lambda} S \bar{N} U \bar{U}+\frac{1}{\Lambda} S \bar{N}^3
\end{equation}
One can check that these operators are more than sufficiently suppressed by a Planck scale cutoff and an order one coupling to avoid excessive proton decay for our chosen reference parameters of $\tan \beta = 10$, $m_{\rm soft} = 300 \mathrm{\ GeV}$, $w''=1$, $\langle \Phi \rangle \sim 10^6 \mathrm{\ TeV}$ and $\langle \phi \rangle \lsim 10^3 \mathrm{\ TeV}$. Dimension six operators can also be significant if they contain at least three flavor Higgs fields. The most significant of these operators are
\begin{equation}
W_{\rm LNV}^{(6)} = \frac{1}{\Lambda^2} \bar{N} \bar{\Phi}_e \bar{\Phi}_u^3+\frac{1}{\Lambda^2} \bar{N} \bar{\Phi}_u \Phi_d^3 + \ldots
\end{equation}
where the omitted terms generate subleading contributions to the $\bar{N}$ tadpole. One can show that these operators are also sufficiently suppressed for $\langle \Phi \rangle \sim 10^6 \mathrm{\ TeV}$.

Planck suppressed operators can also contribute to the electroweak Higgs potential. In particular, we find the dimension-five contributions to the $S$ tadpole:
\begin{equation}
W_{\rm EW}^{(5)} = \frac{1}{\Lambda} S \phi_d \Phi_d \bar{\Phi}_e + \frac{1}{\Lambda} S \phi_d \Phi_e \bar{\Phi}_u
\end{equation}
However, one can check that these generate a tadpole of only about $(300\mathrm{\ GeV})^2$ for $\langle \Phi \rangle \sim 10^6 \mathrm{\ TeV}$ and $\langle \phi_d \rangle \sim 10^3 \mathrm{\ TeV}$ and thus will not cause a fine tuning of the electroweak scale, and can in fact facilitate electroweak symmetry breaking even in the absence of the SUSY breaking terms.

\section{SUSY breaking and particle spectrum beyond the MSSM} \label{sec:SUSYbreaking}

In this section, we discuss supersymmetry breaking and its consequences for the low energy spectrum, as well as the possible effects of a light right-handed vector-like generation of quarks, such as can occur in our model.

We consider a supersymmetry breaking spurion $X$, a chiral superfield with an F-term vev $\langle X \rangle_F \sim F$, which couples to our model via higher-dimensional operators suppressed by a messenger scale $M$, such that $m_{\rm soft} \sim F/M$. In particular, we focus on the case of gravity mediation, where $M$ is the Planck scale and $X$ may be thought of as a hidden-sector field which couples to the SM sector via Planck-suppressed operators. We will show that, contrary to the usual situation where gravity mediation induces a flavor problem, the gauged flavor symmetry together with the $\bZ_{11}$ gauged R-symmetry will protect against FCNCs, giving an MFV structure at leading order with corrections suppressed by $\langle \Phi \rangle / M \sim 10^{-10}$. Indeed, in this context gravity mediation is actually preferred, as lowering the messenger scale will eventually lead to subleading non-MFV corrections as the messenger scale approaches the flavor scale.

The soft SUSY-breaking squark masses for the right-handed up-type squarks are generated by the effective K\"ahler potential:
\begin{equation}
\int d^4\theta \left[ \frac{X^\dagger X}{M^2} ( a_1 \bar{u}^\dagger \bar{u} +a_2 \bar{U}^\dagger \bar{U})\right] \,,
\end{equation}
where both terms are $\SU(3)_F$ universal due to the gauging of the flavor symmetry. Integrating out the heavy fields, we obtain the soft masses
\begin{equation}
\mathcal{L}_{\rm soft} \supset m_{\rm soft}^2 \tilde{\bar{u}}^{\star} \left( a_1 \mathbf{1}+\frac{a_2-a_1}{|\lambda_u|^2} Y_u^{\dag} Y_u \right) \tilde{\bar{u}} \,,
\end{equation}
and likewise for the right-handed down-type squarks. Thus, the soft terms are MFV to leading order, though they are already non-universal in the high scale theory, even before accounting for the running between the flavor scale and the electroweak scale. Non-MFV corrections will arise from higher-dimensional operators involving the flavor Higgs fields, and will therefore be strongly suppressed for a Planck-scale cutoff.

At first glance, the left-handed squark mass matrix appears to be universal at the flavor scale, arising from the effective K\"ahler potential
\begin{equation}
\int d^4\theta \left[ \frac{X^\dagger X}{M^2} b_1 q q^{\dag}\right] \,.
\end{equation}
However, there are potentially important corrections upon integrating out the heavy vector-like generations coming from the effective K\"ahler potential
\begin{equation}
\int d^4\theta \left[ \frac{X^\dagger X}{M^2} (b_2 U^{\dag} U + b_3 D^{\dag} D)\right] \,.
\end{equation}
Upon integrating out the heavy fields we obtain the squark masses
\begin{multline}
b_1 m_{\rm soft}^2 \tilde{q} \tilde{q}^{\star} + b_2 m_{\rm soft}^2 \frac{v_u^2}{|\mu_u|^2} \tilde{u}_L Y_u Y_u^{\dag} \left(1-\frac{1}{|\lambda_u|^2} Y_u Y_u^{\dag} \right) \tilde{u}_L^{\star}\\+ b_3 m_{\rm soft}^2 \frac{v_d^2}{|\mu_d|^2} \tilde{d}_L Y_d Y_d^{\dag} \left(1-\frac{1}{|\lambda_d|^2} Y_d Y_d^{\dag} \right) \tilde{d}_L^{\star} \,,
\end{multline}
so there are tree-level non-universal MFV contributions to the squared mass matrix suppressed by $(v/\mu)^2$, in addition to the usual RG corrections.

The soft breaking A-terms will be holomorphic MFV to leading order. For example the effect of the $\bar{U}\bar{D}\bar{D}$ operator is
\begin{equation}
c_1 \int d^2\theta \frac{X}{M} \bar{U}\bar{D}\bar{D} \ \to \  c_1 \frac{m_{\rm soft}}{\lambda_u \lambda_d^2} (Y_u \tilde{\bar{u}}) (Y_d \tilde{\bar{d}})(Y_d \tilde{\bar{d}}) \, .
\end{equation}
Similarly, the A-terms corresponding to the ordinary Yukawa couplings are 
 \begin{equation}
c_2 \int d^2\theta \frac{X}{M} q \bar{U} H_u \ \to \  c_2 \frac{m_{\rm soft}}{\lambda_u}  (\tilde{q} \hat{H}_u) Y_u \tilde{\bar{u}}  \, ,
\end{equation}
where $\hat{H}_u$ denotes the scalar component of $H_u$. Certain non-holomorphic combinations of spurions can also appear in the A-terms:
 \begin{multline} \label{eq:nonholAterm}
c_3 \int d^2\theta \frac{X}{M} \phi U \bar{u} \ \to \ c_3 m_{\rm soft} \frac{\langle\phi\rangle}{\mu_u} (\tilde{q} \hat{H}_u) \left(1-\frac{1}{|\lambda_u|^2} Y_u Y_u^{\dag}\right) Y_u \tilde{\bar{u}}  \\ +c_3 m_{\rm soft} \frac{w'' \langle\phi\rangle}{2 \mu_u} \left[\left(1-\frac{1}{|\lambda_u|^2} Y_u Y_u^{\dag}\right) Y_u \tilde{\bar{u}}\right]\left[Y_d \tilde{\bar{d}}\right]^2 \, ,
\end{multline}
and likewise for the $\phi D \bar{d}$ A-term. Note that $\langle\phi\rangle/\mu\sim1$, so these are non-holomorphic MFV corrections with order one coefficients, but they take a very particular form which was anticipated already in~\cite{MFVSUSY} and which is not in any way problematic.

However, there are additional sources of A-terms --- some of which may be dangerous --- from SUSY breaking terms of the form
\begin{equation} c_4 \int d^2\theta \frac{X}{M} U \Phi \bar{U} \,.
\end{equation}
Upon integrating out the heavy fields as usual we find 
\begin{equation}
{\mathcal L}_{soft}\supset  c_4 \frac{m_{\rm soft}}{\mu_u \lambda_u} (\tilde{q} \hat{H}_u) \sqrt{1-\frac{Y_u Y_u^{\dag}}{|\lambda_u|^2}} Y_u \langle \Phi \rangle Y_u \tilde{\bar{u}} + c_4 w'' (\ldots)\, .
\end{equation}
If $\langle \Phi_u \rangle \propto {\mathcal M}_u$ then we get an additional MFV contribution of the same form as~(\ref{eq:nonholAterm}). However, in the model based on the gauged $\bZ_{11}$ R-symmetry both $\Phi_u$ and $\Phi_n$ carry the same charges. This would not be problematic if the same linear combination of these fields were to appear in both the superpotential and the soft-terms, but there is no a priori reason for this to occur unless enforced by some symmetry principle. Conversely, if both combinations are allowed in the A-terms, then that A-term would contain a additional structure proportional to $Y_u M_N Y_u$ which deviates from the MFV form by an essentially arbitrary $3\times3$ symmetric matrix, contributing to off-diagonal holomorphic non-MFV squark masses (though still Yukawa suppressed). In order to forbid such contributions, one can for example introduce an additional $\bZ_2$ discrete gauge symmetry, under which $\Phi_{u,d}, \bar{\Phi}_{u,d}$ and $U,D$ are odd, and every other field is even. This $\bZ_2$ is also anomaly free, and forbids the mixing of the $\Phi_u$ and $\Phi_n$ fields, but will also restrict the form of the general Higgs potential of (\ref{eq:HiggsPotential}). To avoid this problem, one can for instance introduce two copies of each $\Phi$ and $\phi$ Higgs field variant labeled $\Phi^{\pm}$, such that the $+$ and $-$ Higgs fields are respectively even and odd under the $\bZ_2$. Thus, $\Phi_u \equiv \Phi_u^{-}$ will generate the up-sector Yukawa couplings, whereas $\Phi_n \equiv \Phi_u^{+}$ will generate the neutrino masses, and no mixing between the two is permitted by the $\bZ_2$. (This extension of the Higgs sector allows a richer Higgs potential, which may in any case be needed to obtain the desired flavor structure.) Another possible solution is to choose the flipped embedding of $\SU(3)_F \subset \SU(3)_Q \times \SU(3)_L$, so that the quark flavor structure is generated by $\bar{\Phi}$'s whereas that of the leptons is generated by $\Phi$'s.
 
Finally, we address the question of whether any of the additional particles in our model (beyond the NMSSM) could be within reach of the LHC and what their signals could be. As discussed in section~\ref{sec:FCNC}, the constraints from FCNC's will force the flavor gauge bosons to be at $10^4-10^6$ TeV, well outside the LHC's range. Similarly, most of the heavy vectorlike quarks $U,\bar{U},D,\bar{D}$ will be too heavy for LHC energies, since their masses are determined by the same flavor Higgs VEVs $\Phi_{u,d}$ that contribute to the flavor gauge boson masses. However, in order for the top quark to have an ${\cal O}(1)$ Yukawa coupling, the corresonding $U,\bar{U}$ should have one eigenvalue ${\mathcal M}_u^{(3)}$which shold be comparable or smaller than the corresponding mixing term $\mu_u$, which cannot itself exceed about $10\mathrm{\ TeV}$ in order to generate the large up/top hierarchy if the flavor scale is $10^6\mathrm{\ TeV}$. These parameters are not strongly constrained by FCNC's, and could lie within the LHC energies. To study the phenomenology of the third generation up-type quarks we focus on their interactions, neglecting the other generations:
\begin{equation}
{\cal L} \supset \mu_u t_R^1 t_L^2 + {\mathcal M}_u t_R^2 t_L^2 +\lambda _u q^{(3)} t_R^1 H_u \ ,
\end{equation}
where we introduced the notation used in the little Higgs literature for top partners $\bar{u}^{(3)} = t_R^1, U^{(3)} =t_L^2, \bar{U}^{(3)} = t_R^2, q^{(3)}= (t_L^1, b_L)$ and ${\mathcal M}_u = {\mathcal M}_u^{(3)}$. The mass of the heavy vector partners is given by 
\begin{equation}
m_T = \sqrt{\mu_u^2+{\mathcal M}_u^2} \ ,
\end{equation}
 with the mixing among the right handed quarks is given by the angle 
\begin{equation}
\sin \alpha = \frac{\mu_u}{\sqrt{\mu^2_u+{\mathcal M}_u^2}}\ .
\end{equation}
The top quark mass is given by 
\begin{equation}
m_t= \lambda_u \cos \alpha \frac{v_u}{\sqrt{2}} \ .
\end{equation}
A mixing among the left handed top quarks is induced after EWSB and is given by  the mixing angle
\begin{equation}
\sin \gamma = \frac{\lambda_u \mu_u v_u/\sqrt{2}}{\mu_u^2 +{\mathcal M}_u^2}= \frac{m_t}{m_T} \tan \alpha\ .
\end{equation}
The mixing pattern is the same as for the heavy top partners in little Higgs models, and this will largely detrmine the phenomenology of these models. The main difference is that in our case the cancellation of the quadratic divergences is achieved via SUSY, rather than through the non-linearly realized SU(3) symmetry of the little Higgs models. However, this does not affect the phenomenology of the top partners. The couplings of the top partners to gauge bosons is discussed in detail in Appendix A of~\cite{Maxim}. Electroweak precision correction bounds from loops of the heavy vectorlike top partners is around 450 GeV as long as the mixing angle $\alpha$ is not too small~\cite{Maxim}. The direct production bounds from the 2011 dataset of 5 $fb^{-1}$ is somewhat weaker, of order 350 GeV, while a more recent analysis puts a more stringent direct bound of 480 GeV on the mass of the top partners~\cite{newtoppartnersearch}.\footnote{The superpartners of the top partners would just behave like heavy stops: their pair production cross section is very small, and they would then decay to the LSP and finally through the RPV term to jets.} Thus we conclude that the third generation $U,\bar{U}$ states can be below 1 TeV and within the range of the 14 TeV LHC, but this need not be the case: they can be as heavy as $10\mathrm{\ TeV}$ for ${\mathcal M}_u^{(1)} \sim 10^6$ TeV. 

\section{Conclusions} \label{sec:conclusions}

We have presented a complete model which violates baryon number and R-parity in a controlled fashion, leading to prompt LSP decay and low energy energy signatures which evade the stringent LHC bounds on R-parity conserving supersymmetry broken at the electroweak scale. At the same time, our model solves the $\mu$ problem as well as the flavor problem of gravity mediated supersymmetry breaking, provides a potential explanation for the origin of flavor in the standard model, and is safe from Planck suppressed corrections.

We accomplish this by gauging an $\SU(3)_F$ flavor symmetry at high energies and spontaneously breaking it. After integrating out the massive fields (vector-like right handed generations) the universal Yukawa couplings and $\bar{U}\bar{D}\bar{D}$ BNV coupling are simultaneously reduced to the low-energy hierarchical Yukawa couplings and a $\bar{u}\bar{d}\bar{d}$ R-parity violating BNV coupling of the MFV SUSY form. We introduce a gauged discrete R-symmetry to forbid other sources of baryon number violation as well as excessive lepton number violation. This discrete symmetry also allows us to solve the $\mu$ problem by introducing NMSSM singlet(s) $S$ and forbidding the super-renormalizable terms in the Higgs potential via the discrete symmetry. We exhibit an example of a $\bZ_{11}$ discrete R-symmetry which accomplishes all of these goals while allowing a suitably rich potential for the flavor Higgs sector and protecting the model from dangerous Planck-suppressed corrections.

The gauged $\SU(3)_F$ symmetry ensures that soft SUSY breaking terms are MFV to leading order, but with a non-universal structure which allows for flexibility in the low-energy superpartner spectrum. As FCNC constraints require a flavor scale of at about $10^6\mathrm{\ TeV}$ or higher, the flavor gauge bosons will be out of reach of the LHC. However, the third generation of right-handed vector-like up-type quarks must be much lighter than the flavor scale to generate the large up/top mass hierarchy, and could lie below $1\mathrm{\ TeV}$. In this case it would have collider properties similar to the top partners in little Higgs models. 

Since we have gauged only a single $\SU(3)_F$ for both quarks and leptons, this type of model may be embeddable in an $\SU(5)$-type GUT. We explore this possibilitiy in a future work~\cite{future}.

\section*{Acknowledgements}

While we were concluding this project, two papers with similar models have been published~\cite{Krnjaic:2012aj,Roberto}. We thank Gordan Krnjaic for informing us ahead of time of the release of their paper and for useful discussions related to this work. We also thank Maxim Perelstein for useful discussions.  This research was supported in part by the NSF grant PHY-0757868.

\section*{Appendix}

\appendix

\section{Choosing the discrete symmetry\label{sec:appendix}}

In this appendix, we search for an anomaly-free discrete symmetry which allows all of the necessary terms in the superpotential~(\ref{eqn:WEEbar}) or~(\ref{eqn:WLLbar}) while forbidding all of the problematic operators, (\ref{eq:badBNV}), (\ref{eq:badUnivDir})--(\ref{eq:badFlipDir}), (\ref{eq:badUnivL})--(\ref{eq:badStdQQQL}), (\ref{eq:badEW1}), and (\ref{eq:badEW2}).

In particular, for simplicity we focus on the model with $E,\bar{E}$ leptons and the standard embedding of $\SU(3)_F$ within $\SU(3)_Q\times\SU(3)_L$. Requiring that the superpotential~(\ref{eqn:WEEbar}) transforms as $W \to \eta_W W$, we find that an arbitrary discrete symmetry of the theory takes the form:
\begin{center}
  \begin{tabular}{|c|ccccc|}
\hline
    & $\SU(3)_C$ & $\SU(2)_L$ & $\U(1)_Y$ & $\SU(3)_F$ & $\bZ_k  [ \eta_S^3]$  \\ \hline
    $q$ & $\fund$ & $\fund$ & $1 / 6$ & $\fund$ & $\eta_S$ \\
    $\bar{u}$ & $\ov{\fund}$ & $\sing$ & $- 2 / 3$ & $\fund$ &  $\eta_{\bar{u}}$ \\
    $\bar{d}$ & $\ov{\fund}$ & $\sing$ & $1 / 3$ & $\fund$ &  $\eta_{\bar{d}}$ \\
    $\ell$ & $\sing$ & $\fund$ & $- 1 / 2$ & $\fund$ & $\eta_S^2 \eta_{\bar{E}}^{- 1}$ \\
    $\bar{e}$ & $\sing$ & $\sing$ & $1$ & $\fund$ &  $\eta_{\bar{e}}$ \\
    $\bar{U}$ & $\ov{\fund}$ & $\sing$ & $- 2 / 3$ & $\ov\fund$ & $\eta_S$  \\
    $\bar{D}$ & $\ov{\fund}$ & $\sing$ & $1 / 3$ & $\ov\fund$ & $\eta_S$ \\
    $\bar{E}$ & $\sing$ & $\sing$ & $1$ & $\ov\fund$ & $\eta_{\bar{E}}$ \\
    $U$ & $\fund$ & $\sing$ & $2 / 3$ & $\ov\fund$ & $\eta_U$ \\
    $D$ & $\fund$ & $\sing$ & $- 1 / 3$ & $\ov\fund$ & $\eta_D$ \\
    $E$ & $\sing$ & $\sing$ & $- 1$ & $\ov\fund$ & $\eta_E$ \\
    $\bar{N}$ & $\sing$ & $\sing$ & $0$ & $\ov\fund$ & $\eta_{\bar{E}}$ \\
    $H_u$ & $\sing$ & $\fund$ & $1 / 2$ & $1$ & $\eta_S$ \\
    $H_d$ & $\sing$ & $\fund$ & $- 1 / 2$ & $1$ & $\eta_S$ \\
    $S$ & $\sing$ & $\sing$ & $0$ & $1$ & $\eta_S$ \\ \hline
  \end{tabular}
\end{center}
in the ``matter'' sector after mixing with an arbitrary subgroup of $\U(1)_Y$ and the $\bZ_3$ center of $\SU(3)_C$, where $\bZ_k [ \eta_W ]$ denotes a discrete R-symmetry under which $W \to \eta_W W$ (i.e. $\theta \to \eta_{\theta} \theta$ where $\eta_W = \eta_{\theta}^2$). One can then read off the action of the discrete symmetry on the flavor Higgs sector:
\begin{center}
  \begin{tabular}{|c|cc|}
\hline
   &  $\SU(3)_F$ & $\bZ_k  [ \eta_S^3]$  \\ \hline
    $\Phi_u$ &   $\symm$ & $\eta_S^2 \eta_U^{- 1}$  \\
    $\Phi_d$ &   $\symm$ & $\eta_S^2 \eta_D^{- 1}$     \\
    $\Phi_e$ &   $\symm$ & $\eta_S^3 \eta_E^{- 1}\eta_{\bar{E}}^{- 1}$  \\
    $\Phi_n$ &  $\symm$ & $\eta_S^3 \eta_{\bar{E}}^{- 2}$  \\
    $\phi_u$ &   $1$ & $\eta_S^3 \eta_U^{- 1} \eta_{\bar{u}}^{- 1}$ \\
    $\phi_d$ &   $1$ & $\eta_S^3 \eta_D^{- 1} \eta_{\bar{d}}^{- 1}$  \\
    $\phi_e$ &   $1$ & $\eta_S^3 \eta_E^{- 1} \eta_{\bar{e}}^{- 1}$  \\
\hline
\end{tabular}
\end{center}
Henceforward we make the simplifying assumption that the flavor Higgs sector is completely vector-like, i.e.\ that there exist fields $\bar{\Phi}_u$, $\bar{\Phi}_d$, etc.\ such that the mass terms $W_{\rm mass} = M_u \Phi_u \bar{\Phi}_u + M_d \Phi_d \bar{\Phi}_d+\ldots$ can appear in the superpotential. This implies that the flavor Higgs sector makes no net contribution to the anomalies.

Discrete gauge symmetries are far less constrained than continuous gauge symmetries, since they lack cubic anomalies~\cite{Banks:1991xj}. In fact, the only anomalies which must be cancelled for a discrete gauge symmetry are the $G^2 \bZ_k$ and $(\mathrm{grav})^2 \bZ_k$ anomalies for all non-abelian gauge group factors $G$ (precisely those anomalies which relate to gauge and gravitational instantons.) The cancelations of the $\SU(3)_C^2\, \bZ_k$ and $\SU(2)_L^2\, \bZ_k$ anomalies impose the constraints 
\begin{equation} \label{eq:SManomaly}
  \eta_{\bar{u}}^3 \eta_{\bar{d}}^3 \eta_U^3 \eta_D^3  =  \eta_S^{15}, \ \ \ 
  \eta_{\bar{E}}^3  =  \eta_S^2\ .
\end{equation}
Assuming that the flavor Higgs sector is vector-like, the $\SU(3)_F^2\, \bZ_k$ anomaly together with the previous conditions requires
\begin{equation} \label{eq:SU3Fanomaly}
\eta_{\bar{e}} \eta_E  =  \eta_S^5\ .
\end{equation}
Finally, cancellation of the $(\mathrm{grav})^2 \bZ_k$ anomaly together with the previous conditions naively requires
\begin{equation} \label{eq:gravanomaly}
 \eta_S^{N_S - 2} = 1 \,,
\end{equation}
where we now allow for an arbitrary number $N_S$ of NMSSM singlets $S$ and ignore any contribution from other hidden sectors of the theory. Such hidden sectors are inevitably present, however, as a truly complete theory will require a SUSY breaking sector, likely with R-charged gauginos, as well as a supergravity completion with an R-charged gravitino. Thus, while we can solve~(\ref{eq:gravanomaly}) by setting $N_S=2$, the true anomaly constraint will depend on details of the hidden sector, and hence there is no clear constraint on $N_S$. It should be noted, however, that regardless of these details the true $(\mathrm{grav})^2 \bZ_k$ anomaly can usually be cancelled by an appropriate choice of $N_S$.

The anomaly constraints~(\ref{eq:SManomaly}) and~(\ref{eq:SU3Fanomaly}) have no analogous caveats, and must be satisfied if no additional SM charged or flavored fields are added to the model. A general solution to these constraints can be parametrized by
\begin{align}
  \eta_{\bar{E}} &= \alpha^2\,, &
  \eta_S &= \alpha^3  \,, &
  \eta_{\bar{e}} &= \alpha^{15} \eta_E^{- 1} \,, &
  \eta_{\bar{u}} &= \omega_3^p \alpha^9 \eta_U^{- 1} \beta^{- 1} \,, &
  \eta_{\bar{d}} &= \omega_3^p \alpha^6 \eta_D^{- 1} \beta \,, \label{eq:anomparams}
\end{align}
for phase factors $\alpha, \beta$ and an integer $p$, where $\eta_W = \eta_S^3 = \alpha^9$. Thus, $\eta_{\phi_e} = \alpha^{-6}$ and $\eta_{\bar{\phi}_e} = \alpha^{15}$ as a consequence of canceling the $\SU(3)_F^2\, \bZ_k$ anomaly. We wish to forbid the problematic cross couplings between the flavor and electroweak Higgs sectors~(\ref{eq:badEW1}), (\ref{eq:badEW2}), which can lead to fine tuning of the electroweak scale. In particular, to forbid $\phi^2 S$ for $\phi \in \{\phi_e, \bar{\phi}_e\}$ we must require
\begin{equation} \label{eq:discrconstr}
  \alpha^{18} \neq 1\,,\;\; \alpha^{24} \neq 1\,.
\end{equation}
Thus $\eta_W=\alpha^9 \neq 1$, and we require an R-symmetry.

One can check that these conditions imply that the couplings~(\ref{eq:badEW2}) are also forbidden, as are the remaining cross couplings in~(\ref{eq:badEW1}) involving only $\phi_e$ and $\bar{\phi}_e$. Suppose that $\phi$ is another flavor Higgs singlet in the theory with charge $\eta_{\phi}$ and conjugate field $\bar{\phi}$. To forbid the cross couplings~(\ref{eq:badEW1}) between $\phi,\bar{\phi},\phi_e,\bar{\phi}_e$ and the electroweak Higgs sector (in particular $\phi_i \phi_j S$,) we must require:
\begin{equation} \label{eq:phicons}
\eta_{\phi} \notin \{\alpha^{-9},\alpha^{-3},\pm \alpha^3, \pm \alpha^6, \alpha^{12}, \alpha^{18} \} \,.
\end{equation}
There are analogous constraints on the charge $\eta_{\Phi}$ of a flavor Higgs tensor $\Phi$ with conjugate field $\bar{\Phi}$ in order to forbid the cross couplings~(\ref{eq:badEW1}) as well as the $\bar{N}$ tadpole~(\ref{eq:badUnivL}). Using $\eta_{\Phi_n} = \alpha^5$ and $\eta_{\bar{\Phi}_n} = \alpha^4$, we obtain the constraints
\begin{equation} \label{eq:Phicons}
\eta_{\Phi} \notin \{\alpha^{-4}, \alpha^{-1},\pm \alpha^{1/2}, \alpha^2, \omega_3^{\pm 1} \alpha^2, \alpha^7, \omega_3^{\pm 1} \alpha^7,\pm \alpha^8, \alpha^{11}\} \,.
\end{equation}
These constraints limit the allowed charges of the flavor Higgs fields and hence the form of the Higgs potential. We impose minimum consistency requirements on the flavor Higgs potential: it must contain at least one $\Phi^3$ and at least one $\bar{\Phi}^3$ operator (or else the F-term conditions set the fields to zero) and it must not have any accidental continuous symmetries. Given these requirements, we now search for the smallest possible discrete R-symmetry which allows an acceptable Higgs potential.

The lowest-order $\bZ_k$ R-symmetries that do not contradict~(\ref{eq:discrconstr}) are $k=5,7,10,11,13,\ldots$, where $\bZ_{10} \cong \bZ_5\times\bZ_2$. For $k=5$ and $k=7$ one can check that the constraint~(\ref{eq:Phicons}) is so restrictive that $\eta_{\Phi} = \eta_{\Phi_n}$ necessarily, whereas $\Phi_n^3$ and $\bar{\Phi}_n^3$ are forbidden by~(\ref{eq:discrconstr}). For $k=10$, we can choose either $\alpha = \omega_{10}$ or $\alpha = \omega_5$. In the former instance, we find that $\{\eta_{\Phi}\} \subset \{1,\omega_5^2,-1\}$, where $\{\eta_{\Phi}\}$ is a strict subset since the presence of all three variants will generate the $\bar{N} \Phi \bar{\Phi}^2$ tadpole. Since $\eta_{\Phi_n} = -1$, either $\{\eta_{\Phi}\} \subseteq \{1,-1\}$ or $\{\eta_{\Phi}\} \subseteq \{\omega_5^2,-1\}$, but in either case neither $\Phi^3$ nor $\bar{\Phi}^3$ is permitted. For $\alpha = \omega_5$, we find $\eta_{\Phi} \subseteq \{1,\omega_{10}^3,-1,\omega_{10}^{-3},\omega_{10}^{-1}\}$, but to avoid all $\bar{N} \Phi \bar{\Phi}^2$ tadpoles as well as $\Phi \bar{\Phi} S$ cross-couplings, we can have at most one additional variant of $\Phi$ beyond $\eta_{\Phi_n} = 1$, whereas no such pairing allows both $\Phi^3$ and $\bar{\Phi}^3$ interactions.

The next lowest order possibility is $k=11$, which we will show to be sufficient. We choose $\alpha = \omega_{11}$ so that $\eta_W = \omega_{11}^{- 2}$. The above constraints dictate:
\begin{equation}
  \{ \eta_{\Phi} \} \subseteq \{ \omega_{11}, \omega_{11}^3, \omega_{11}^4, \omega_{11}^5, \omega_{11}^{-2} \} \,.
\end{equation}
However, one can show that to forbid all of the cross couplings~(\ref{eq:badEW1}) and the $\bar{N} \Phi \bar{\Phi}^2$ tadpole we must have $\{ \eta_{\Phi} \} \subseteq \{ \omega_{11}, \omega_{11}^3, \omega_{11}^5\}$, $\{ \eta_{\Phi} \} \subseteq \{ \omega_{11}^4, \omega_{11}^5\}$ or $\{ \eta_{\Phi} \} \subseteq \{\omega_{11}^{-2}, \omega_{11}^5\}$. No $\Phi^3$ or $\bar{\Phi}^3$ interactions are possible in the latter two cases, so we consider the first case. The possible cubic interactions are:
\begin{gather}
\begin{aligned}
  \Phi^3 &:  \omega_{11} \cdot \omega_{11}^3 \cdot \omega_{11}^5 \;, \;
  \omega_{11}^3 \cdot \omega_{11}^3 \cdot \omega_{11}^3 \,,\\
  \bar{\Phi}^3 &: \omega_{11}^8 \cdot \omega_{11}^8 \cdot \omega_{11}^4 \;,
  \; \omega_{11}^8 \cdot \omega_{11}^6 \cdot \omega_{11}^6 \,.
\end{aligned}
\end{gather}
Since $\eta_{\Phi_n} = \omega_{11}^5$, all three $\Phi$ variants must be
present to generate both $\Phi^3$ and $\bar{\Phi}^3$.

We also have
\begin{equation}
  \{ \eta_{\phi_i} \} \subseteq \{1, \omega_{11}^4, \omega_{11}^5, \omega_{11}^{-2}, \omega_{11}^{-1} \} \,.
\end{equation}
Since $\eta_{\phi_e} = \omega_{11}^5$ and $\eta_{\bar{\phi}_e} = \omega_{11}^4$ these variants are always present, and one can show that the Higgs potential has an accidental $\U(1)_R$ symmetry unless $\omega_{11}^{-1} \in \{ \eta_{\phi_i} \}$ as well. The $\{1, \omega_{11}^{-2}\}$ variants are not necessary, but neither are they problematic. We will assume that they are absent for simplicity. Assuming $\{ \eta_{\Phi} \} = \{ \omega_{11}, \omega_{11}^3, \omega_{11}^5 \}$ and $\{ \eta_{\phi_i} \} = \{\omega_{11}^4, \omega_{11}^5, \omega_{11}^{-1} \}$, one can show that the flavor Higgs potential is free of accidental $\U(1)$ symmetries, and that all of the cross couplings~(\ref{eq:badEW1}) and all $\bar{N} \Phi \bar{\Phi}^2$ tadpoles are forbidden, as is $(\ell H_u) \Phi^2 \bar{\Phi}$.

Because we have assumed a $\bZ_{11}$ discrete symmetry, we must set $p=0$ in~(\ref{eq:anomparams}). Thus, $\eta_{\phi_u} \eta_{\phi_d} = \alpha^3$ 
and so $\beta = \eta_{\phi_u} \in \{ \omega_{11}^4, \omega_{11}^{- 1} \}$. We must also require $\eta_E \in \{ \omega_{11}^2, \omega_{11}^4, \omega_{11}^6\}$. So far our discussion applies to both the standard and flipped embeddings of $\SU(3)_F \subset \SU(3)_Q\times\SU(3)_L$. We now specialize to the standard embedding, which implies that $\eta_U,\eta_D\in  \{ \omega_{11}, \omega_{11}^3, \omega_{11}^5\}$. For the case $\beta = \omega_{11}^4$, we must have $\eta_U \neq \omega_{11}^3$ and $\eta_D \neq \omega_{11}$ to forbid $\bar{u} \bar{u} \bar{D} \bar{E}$ and $\phi q \ell \bar{d}$ respectively. To forbid $\bar{\Phi} \bar{u} D \bar{E}$ we require $\eta_U \eta_D^{-1} \in \{\omega_{11}^{-4}, \omega_{11}^{-2}, 1\}$ whereas to forbid $\bar{\Phi} U \bar{d} E$ and $\Phi \bar{u} D \bar{e}$ we require $\eta_U \eta_E \eta_D^{-1} \in  \{ \omega_{11}^7, \ldots, \omega_{11}^{10},1\}$. Since $\eta_E \in \{ \omega_{11}^2, \omega_{11}^4, \omega_{11}^6\}$ this implies that $\eta_U \eta_D^{-1} \in \{\omega_{11}^{-4}, \omega_{11}^{-2}\}$ and $\eta_E \neq  \omega_{11}^6$. Therefore $\eta_U = \omega_{11}$, but to forbid $q^2 U E$ we require $\eta_U \eta_E \neq \omega_{11}^3$, so $\eta_E = \omega_{11}^4$ and thus $\eta_U \eta_D^{-1} = \omega_{11}^{-4}$ so that $\omega_D = \omega_{11}^5$. This is the solution presented in section~\ref{sec:completemodel}.

We also consider the case $\beta = \omega_{11}^{-1}$, so that $\eta_D \neq \omega_{11}^3$ to forbid $\phi q \ell \bar{d}$, whereas $\eta_U \eta_D^{-1} \in \{\omega_{11}^2, \omega_{11}^4\}$ to forbid $\bar{\Phi} \bar{u} D \bar{E}$ so that $\eta_D = \omega_{11}$. To forbid $\bar{\Phi} U \bar{d} E$ and $\Phi \bar{u} D \bar{e}$ we now require $\eta_U \eta_E \in \{ \omega_{11}^2, \omega_{11}^3, \ldots,\omega_{11}^6\}$. However, the only possible solution is $\eta_U \eta_E = \omega_{11}^5$, i.e.\ $\eta_U = \omega_{11}^3$ and $\eta_E = \omega_{11}^2$.

We will not discuss this second model in detail, nor attempt to classify $\bZ_{11}$ models with a flipped embedding. Note than we have not considered discrete gauge symmetries that are irreducible products, e.g. $\bZ_5\times\bZ_5$. It would be interesting to determine if a ``simpler'' discrete gauge symmetry with the right properties can be found in this way.

\end{document}